\documentclass[pra,twocolumn,superscriptaddress,aps]{revtex4}

\usepackage{amsfonts}
\usepackage{amssymb,latexsym,amsmath}
\usepackage{graphicx}
\usepackage{times}

\newcommand{\LC}{\left(}
\newcommand{\RC}{\right)}
\newcommand{\LB}{\left[}
\newcommand{\RB}{\right]}
\newcommand{\LV}{\left|}
\newcommand{\RV}{\right|}
\newcommand{\p}{\partial}
\newcommand{\curl}{\operatorname{curl}}
\newcommand{\dv}{\operatorname{div}}

\graphicspath{{fig-ps/}}

\begin{document}

\title{Exploring Vortex 
Dynamics in the Presence of Dissipation: 
Analytical and Numerical Results
}

\author{D. Yan}
\affiliation{Department of Mathematics and Statistics, University of Massachusetts,
Amherst, MA 01003-4515, USA}

\author{R.\ Carretero-Gonz\'{a}lez}
\affiliation{Nonlinear Dynamical Systems Group\footnote{{\tt URL:} http://nlds.sdsu.edu/},
Computational Science Research Center, and
Department of Mathematics and Statistics,
San Diego State University, San Diego,
CA 92182-7720, USA}

\author{D. J. Frantzeskakis}
\affiliation{Department of Physics, University of Athens, Panepistimiopolis, Zografos, Athens 15784, Greece}

\author{P. G. Kevrekidis}
\affiliation{Department of Mathematics and Statistics, University of Massachusetts,
Amherst, MA 01003-4515, USA}

\author{N. P.\ Proukakis}
\affiliation{Joint Quantum Centre (JQC) Durham-Newcastle, School of Mathematics and Statistics, Newcastle University,
Newcastle upon Tyne, NE1 7RU, United Kingdom}

\author{D. Spirn}
\affiliation{School of Mathematics, University of Minnesota,
Minneapolis, MN 55455, USA}

\begin{abstract}
In this paper, we examine the dynamical properties
of vortices in atomic Bose-Einstein condensates in the presence of phenomenological dissipation,
used as a basic model for the effect of finite temperatures.
In the context of this so-called dissipative Gross-Pitaevskii model, we derive analytical results
for the motion of single vortices and, importantly, for vortex dipoles which have become very relevant experimentally.
Our analytical results are shown to compare favorably to the full numerical solution of the dissipative Gross-Pitaevskii equation in parameter
regimes of experimental relevance. We also present
results on the stability of vortices and vortex dipoles,
revealing good agreement between numerical and analytical results for
the internal excitation eigenfrequencies, which extends even beyond the regime of validity of this equation for cold atoms.
\end{abstract}

\maketitle

\section{Introduction}

Much of the recent work regarding the emerging topic of nonlinear
phenomena in Bose-Einstein condensates (BECs) has revolved
around the theme of vortices. The latter constitute the prototypical
two-dimensional (or quasi-two-dimensional) excitation that arises
in BECs and bears a topological charge. The volume of work on BECs
focusing on the theme of vortices can be well appreciated from the
existence of many
reviews on the
subject~\cite{fetter,fetter2,newton_review,our_review,rcg:BEC_BOOK}.
Part of the fascination with vortices bears its
roots in the significant
connections of this field with other areas of Physics, including
hydrodynamics~\cite{nvortex_probl},
superfluids~\cite{Pismen1999}, and nonlinear optics~\cite{yuri1,yuri2}.

While
vortices~\cite{chap01:vort2,chap01:vort3,foot} and even
robust lattices thereof
\cite{chap01:latt1,chap01:latt2,chap01:latt3} were observed shortly after the experimental realization of BECs,
recent years have shown a considerable increase in the
interest in vortex dynamics. This is in good measure due to the
activity of many
experimental groups.
Among them, we highlight the pioneering work of Ref.~\cite{BPA_KZ}
enabling the
formation of vortices through the so-called Kibble-Zurek
mechanism~\cite{kibble,zurek2};
the latter was originally proposed
for the formation of large scale structure in the universe, by means
of a quench through a phase transition. In Ref.~\cite{BPA_KZ},
the quench
through the phase transition led
to the BEC formation and hence to the spontaneous trapping
of phase gradients and emergence of vortex excitations.
In many of these experiments, not only single vortices, but
also vortex dipoles were observed. This is where another remarkable
contribution came to play~\cite{freilich}.
By pumping a small fraction of the atoms to a different
(unconfined by the magnetic trap) hyperfine state,
it is possible
to extract atoms (e.g.~$\approx 5$\% of the BEC) and image them,
enabling for the first time (in BEC) a minimally destructive
unveiling of the true vortex dynamics {\it as it happens}.
This work and the follow-up efforts of Ref.~\cite{pra_11}, as well as
yet another remarkable experiment~\cite{BPA_10}
revealed the key dynamical relevance of multi-vortex clusters in the form
of {\it vortex dipoles} (VDs). In the experiment of Ref.~\cite{BPA_10}, the
superfluid analogue of flow past a cylinder
was realized, leading to the spontaneous
emergence of such VDs. In Ref.~\cite{pra_11}, the full (integrable at the level
of two vortices) dynamics of the VD case was revealed. The theme
of few-vortex crystals has garnered interest in the case of higher
number-of-vortices variants in the works of Ref.~\cite{bagnato} (3 vortices)
and in the
very recent co-rotating (same charge) vortex case of Ref.~\cite{corot} (3 as well as
4 same charge vortices).

On the other hand, another topic of increasing attention has concerned
the role of finite-temperature induced
 damping of the BEC \cite{Proukakis_Book} that, in turn, leads to anti-damping in the
motion of the coherent structures therein.
In particular, the simpler case of
dark solitons in single-component BECs
has been studied at some length \cite{shl1,shl2,us,ft1,ft2,ft3,ft4,gk,ashton}. In this context, the work
of Ref.~\cite{shl1} (see also Ref.~\cite{shl2}) provided originally a kinetic-equation
along with a study of the Bogolyubov-de Gennes (BdG) equations. There, it
was argued that the dark soliton obeys an anti-damped harmonic oscillator
equation, resulting into trajectories of growing oscillating amplitudes around the center
of the trap confining the BEC:
the first experiments~\cite{han1,han2,nist}
observed the motion of a dark soliton created by phase imprinting at the center of the trap
towards the edge of the trap~\cite{nist}; 
one/more full soliton oscillations were subsequently demonstrated in two distinct experiments~\cite{bongs,heidelberg}
(see also related theoretical work of Refs.~\cite{ba00,freq}).
More recently, the effect of anti-damping was
actually directly observed for the first time ~\cite{Zwierlein} in the related context of dark soliton
oscillations in the unitary Fermi gas.

Anti-damped dynamical equations for dark solitons (in the context of bosons) were derived in Ref.~\cite{us}
by means of a Hamiltonian perturbation approach~\cite{djf} applied
to the dissipative variant of the Gross-Pitaevskii equation (DGPE).
The DGPE was originally introduced phenomenologically by Pitaevskii~\cite{lp}
as a simplified means of accounting (through a damping term) for the
role of finite temperature induced fluctuations in the condensate
dynamics; see, e.g., Refs.~\cite{penckw,Proukakis_Book,npprev,blakie,ZNG_Book} for the microscopic
interpretation of such a term. It should be noted here
that, at least in the dark soliton context, the DGPE model was found \cite{us} to yield predictions
that compared favorably with more complex stochastic Gross-Pitaevskii
equation (SGPE)~\cite{stoof_sgpe,stoch} {\it when comparing to appropriately averaged quantities with
the same dissipation parameter};
for this reason, we adopt the DGPE in
what follows. The DGPE has been employed in more complex
settings involving multiple dark solitons~\cite{dcdss} and
dark-bright solitons in multiple component BECs~\cite{njp}, yet it has arguably received
somewhat less attention in higher dimensions, and especially
so in the case of single- and multiple vortex states.
The dynamics of vortices under the influence of
thermal effects, has been considered in some computational
detail recently~\cite{proukv1,proukv2,ourjpb,ashtone,tmwright}, and a semi-classical
limit case in the absence of a trap has been of mathematical interest
as well~\cite{spirn,miot} (see also Ref.~\cite{stamp} for the role of quantum effects,
mostly relevant at extremely low temperatures, as well as atom numbers,
where they dominate over thermal effects). 
While the case of a single trapped
vortex has been considered in the pioneering work of Ref.~\cite{stoof}
(see also Ref.~\cite{ashtone2}), 
a comparison of numerical
computations to analytical results, and more importantly, its generalization
to
multi-vortex settings is still an open problem; 
it is the aim of the present work to contribute towards this direction.


We start by developing in Sec.~II an energy-based method that provides a dynamical
equation characterizing the single vortex dynamics,
with our results also supplemented by another technique (cf.~Appendix~\ref{appA})
based on methods similar to the ones of Refs.~\cite{spirn,miot}.
We believe that our method provides a useful alternative
to the approach of Ref.~\cite{stoof}
(the relevant connections are discussed herein).
The main focus of this paper (Sec.~III) is to generalize this to the experimentally
relevant case of a vortex dipole, using the relevant
set of first-order ordinary differential equations (ODEs).
The resulting ODEs are then compared to the
full DGPE model.
We also perform a linear stability analysis
in the absence/presence of the
phenomenological damping,
and characterize the internal modes of the
single and dipole vortex systems (cf.~Appendix~\ref{appB}).
Our results for the single vortex and vortex dipole stability and dynamics reveal the following:
already in the regime of small but experimentally-relevant chemical
potentials, 
where the vortices can be well approximated as particles fully
characterized by their position, the model
provides a qualitatively accurate and semi-quantitative approximation of the relevant stability
and dynamical properties of the original equation (DGPE) for 
typical low temperature BEC experiments.
We summarize and discuss the broader context of our results in Sec.~IV.

\section{Single Vortex Solutions of the Dissipative Gross-Pitaevskii Equation}

The dissipative GPE model can be expressed in the following dimensionless
form for a pancake shaped BEC (see e.g., Ref.~\cite{ourjpb} for the reductions
that lead to such a quasi-two-dimensional description)
\begin{eqnarray}
(i-\gamma) u_t &=& -\frac{1}{2}\Delta u+V(r)u+|u|^2u-\mu u.
\label{dissGPE}
\end{eqnarray}
Here, $u$ represents the quasi-two-dimensional BEC
wavefunction, $\Delta$ represents the Laplacian operator,
$V(r)=\frac{1}{2}\Omega^2r^2$
is the external harmonic trap ($\Omega$ is the normalized
trap frequency, while $r^2=x^2+y^2$), and $\mu$ stands for the chemical potential.
The latter is directly related to the number of atoms in the BEC, with the
limit of large $\mu$ being suitable for a particle-based description of
coherent structures (in a semi-classical fashion; see e.g., Ref.~\cite{coles}).
Finally,
the dimensionless parameter $\gamma$ is associated with the system's
temperature, based on the
earlier treatment of Ref.~\cite{penckw} ---see also
Refs.~\cite{Proukakis_Book,npprev,blakie,ZNG_Book,stoof} for more details.
Physically relevant cases correspond to $\gamma\ll1$ \cite{footnote1},
as also discussed in the specific applications of
Refs.~\cite{us,ft1,ft3}.
%
%

In the case of a single vortex, the stationary state is well-known
to exist at the
center of the parabolic trap; in the Hamiltonian
case of $\gamma=0$, upon displacement from the trap center, the
vortex executes a circular precession around it~\cite{fetter} which
has been also very accurately quantified experimentally~\cite{freilich}.
However, for $\gamma \neq 0$, the anomalous (internal) mode of the
vortex associated with the precessional motion becomes {\it unstable}.
This anomalous mode, as explained, e.g., in Ref.~\cite{ourjpb}, is associated
with the fact that the vortex does not represent the ground state of
the system. Nevertheless, it is a dynamically {\it stable} entity
in the absence of a dissipative channel, as it cannot shed energy
away to spontaneously turn into the ground state.
Yet, as was rigorously
proved in Ref.~\cite{sandst}, the existence of any dissipation
typically renders such anomalous modes (so-called negative
Krein sign modes) {\it immediately} unstable, as it provides
a channel enabling the expulsion of the coherent structure and
the conversion of the system into its corresponding ground state.
This is evident from the presence of a positive imaginary part of the
eigenfrequency (i.e., a real part of the corresponding eigenvalue), which directly
signals the relevant instability for any non-zero value of $\gamma$.
This is discussed further in Appendix~\ref{appB}, which also shows how
the spectrum of the Hamiltonian ($\gamma=0$) case gets modified by $\gamma$.

Motivated by the above discussion (and corresponding
numerical results in Appendix~\ref{appB}), we now
develop a systematic approach based on the time evolution of the vortex
energy for accounting the effect of the anti-damping term
proportional to $\gamma$.
In particular, we will seek to identify
the equations of motion for the center of a single vortex $(x_0,y_0)$
as a first step towards investigating the vortex dipole case.

In the case of $\gamma=0$, the energy of the system [i.e., the Hamiltonian associated with
Eq.~(\ref{dissGPE})] reads:
\begin{equation}
E=\frac{1}{2}
\iint\left\{ \left(|u_x|^2+|u_y|^2\right)+(|u|^2\!-\!\mu)^2+2V|u|^2 \right\}dx\,dy.
\end{equation}
%
%
It is then straightforward to confirm by means of direct calculation
that, in the case of $\gamma \neq 0$,
%
\begin{equation}
\frac{dE}{dt}=-2\gamma\iint|u_t|^2 dx\,dy,
\label{energy_conservation_gamma}
\end{equation}
which forms the starting point of our analysis.

\subsection{Single Vortex Dynamics}

For the case of a single vortex inside the BEC, it can be found
\cite{lundh} that the energy reads:
\begin{align}
\nonumber
E =& \frac{\pi\rho_0}{2}\left[\left(1-\frac{r_0^2}{R_{\rm TF}^2}\right)
\log\left(\frac{R_{\rm TF}^2}{\xi_0^2}\right)\right.\\
&+\left.\left(\frac{R_{\rm TF}^2}{r_0^2}+1-2\frac{r_0^2}{R_{\rm TF}^2}\right)
\log\left(1-\frac{r_0^2}{R_{\rm TF}^2}\right)\right],
\label{E}
\end{align}
where $r_0=\sqrt{x_0^2+y_0^2}$ and $(x_0,y_0)$ is the location of the vortex center,
$R_{\rm TF}=\sqrt{2\mu}/\Omega$
is the Thomas-Fermi radius, $\rho_0=\mu$ and
$\xi_0
=(2\mu)^{-\frac{1}{2}}$ are, respectively, the density and the value of the healing length at the trap center.
For small $r_0/R_{\rm TF}$, i.e., for vortices close to the trap center, the first term in
Eq.~(\ref{E}) dominates and thus:
\begin{equation}
E \approx \frac{\pi\rho_0}{2}\left(1-\frac{r_0^2}{R_{\rm TF}^2}\right)
\log\left(\frac{R_{\rm TF}^2}{\xi_0^2}\right).
\label{Eapprox}
\end{equation}
Then, we have
\begin{eqnarray}
\frac{dE}{dt}
&\approx& \frac{\pi\rho_0}{2}
\left[\log\left(\frac{R_{\rm TF}^2}{\xi_0^2}\right)
\left(-\frac{1}{R_{\rm TF}^2}\right)(2x_0\dot{x}_0+2y_0\dot{y}_0)\right]
\nonumber
\\[1.0ex]
 &\approx& -\omega_{\rm pr}(2x_0\dot{x}_0+2y_0\dot{y}_0)
\end{eqnarray}
where $\omega_{\rm pr}=\frac{\pi\rho_0}{2}
\log(\frac{R_{\rm TF}^2}{\xi_0^2})(\frac{1}{R_{\rm TF}^2})$
is an approximation to the vortex precession frequency.

Now, we turn our attention to the right hand side of
Eq.~(\ref{energy_conservation_gamma}).
Starting from Eq.~(\ref{dissGPE}) with $\gamma=0$ and substituting the following
polar representation of a single vortex $u(r,\theta) = \sqrt{\rho(r)}\,e^{i\theta}$
to Eq.~(\ref{dissGPE}), we obtain the familiar ODE for the
radial vortex profile:
%
%
%
\begin{eqnarray}
\rho''-\frac{\rho'^2}{2\rho}+\frac{\rho'}{r}-\frac{2\rho}{r^2}
+4\left[\mu-\rho-V(r)\right]\rho &=& 0.
\end{eqnarray}
The proposed form for $\rho$ (from a Pad{\'e} approximation ---see
e.g., Ref.~\cite{berloff}) is:
\begin{eqnarray}
\rho(r) &=& \frac{r^2(a_1+a_2r^2)}{1+b_1r^2+b_2r^4}e^{-\Omega^2r^2},
\label{eq:Pade}
\end{eqnarray}
where $a_2=\mu b_2$. The coefficients
$a_1$, $b_1$, and $b_2$ computed by substitution for different choices of
chemical potential $\mu$ considered before, and for a trap frequency of
$\Omega=0.2$, are depicted in Table \ref{tablePade}.

\begin{table}
\begin{tabular}{|c|c|c|c|c|}
\hline
$\mu$ & $a_1$ & $a_2$ & $b_1$ & $b_2$ \\
\hline
0.4 & 0.06698 & 0.01470 & 0.3795 & 0.03676 \\
\hline
0.8 & 0.3676 & 0.05004 & 0.4961 & 0.0625 \\
\hline
1& 0.6021 & 0.0972 & 0.6215 & 0.0972 \\
\hline
1.6 & 1.6359 & 0.4123 & 1.0121 & 0.2577 \\
\hline
\end{tabular}
\caption{
\label{tablePade}
Coefficients  for the Pad\'e approximation (\ref{eq:Pade})
of a unitary charge vortex for different values of the chemical potential $\mu$
for a trap strength of $\Omega=0.2$.
}
\end{table}

A subsequent substitution and direct evaluation
for the right hand side of Eq.~(\ref{energy_conservation_gamma})
reads as follows:
\begin{align}
\nonumber
&-2\gamma\iint|u_t|^2dx\,dy\\[1.0ex]
\nonumber
&\qquad=-2\gamma\iint|u_t(x-x_0(t),y-y_0(t))|^2 dx\,dy\\[1.0ex]
\nonumber
&\qquad=-2\gamma\iint\left(\frac{\partial v}{\partial \xi}(-\dot{x}_0)+\frac{\partial v}{\partial \eta}(-\dot{y}_0)\right)\\[0.5ex]
\nonumber
&\quad\qquad\qquad\qquad\left(\frac{\partial v^{*}}{\partial \xi}(-\dot{x}_0)+\frac{\partial v^{*}}{\partial \eta}(-\dot{y}_0)\right) d\xi\,d\eta\\[1.0ex]
&\qquad=-2\gamma\left(\dot{x}_0^2s+\dot{y}_0^2s\right)
\end{align}
where
\begin{equation}
s=\iint\frac{\partial v}{\partial\xi}\frac{\partial v^{*}}{\partial \xi}=\iint\frac{\partial v}{\partial \eta}\frac{\partial v^{*}}{\partial \eta},
\end{equation}
and ${\rm Re}(\iint\frac{\partial v}{\partial \eta}\frac{\partial v^{*}}{\partial \xi})=0$. Here, we have used the
variables $\xi=x-x_0(t)$ and $\eta=y-y_0(t)$, so that
$v(\xi, \eta)=u(x-x_0(t), y-y_0(t))$.
The resulting integral constant $s$ can be directly evaluated
using the above Pad{\'e} approximation $u=\sqrt{\rho}\,e^{i\theta}$,
finding that
$s=0.5864, 1.5977, 2.1003, 3.5911$ for $\mu=0.4, 0.8, 1, 1.6$ respectively.

Using the above results at hand,
one can combine the left and right hand side of
Eq.~(\ref{energy_conservation_gamma}) and obtain
\begin{equation}
\omega_{\rm pr}(x_0\dot{x}_0+y_0\dot{y}_0) = s\gamma\left(\dot{x}_0^2+\dot{y}_0^2\right).
\label{eq1}
\end{equation}
From this, we can infer the equations of motion of the single vortex state.
Looking for equations of motion that correspond to rotation with anti-damping,
we add a term proportional to $\dot{x}_0\dot{y}_0$ to both sides of Eq.~(\ref{eq1})
and split it into the following two equations:
%
%
%
\begin{eqnarray}
\omega_{\rm pr}x_0+\dot{y}_0 &=& s\gamma\dot{x}_0,
\label{x_dot_revise}
\\[1.0ex]
\omega_{\rm pr}y_0-\dot{x}_0 &=& s\gamma\dot{y}_0.
\label{y_dot_revise}
\end{eqnarray}
%
%
%
Then, the analytical expression for the complex eigenfrequency
$\omega=\omega_r + i\,\omega_i$ is
\begin{eqnarray}
\omega_i&={\rm Im}(\omega)&=\frac{\omega_{\rm pr}s\gamma}{1+(s\gamma)^2},
\label{omega_i}
\\[1.0ex]
\omega_r&={\rm Re}(\omega)&=\frac{\omega_{\rm pr}}{1+(s\gamma)^2}.
\label{omega_r}
\end{eqnarray}
%
%
%
%
%
At this level of approximation, the vortex (outward
spiraling) trajectories can be given explicitly as:
%
\begin{eqnarray}
x &=& e^{\omega_i t}\left[y_0\sin\left(\omega_r t\right)+x_0\cos\left(\omega_r t\right)\right],
\label{solution_x_dissipation}
\\[1.0ex]
y &=& e^{\omega_i t}\left[y_0\cos\left(\omega_r t\right)-x_0\sin\left(\omega_r t\right)\right].
\label{solution_y_dissipatiojn}
\end{eqnarray}
This is a result which parallels the one derived for the case
of the dark soliton (see, e.g., Ref.~\cite{us}). We now provide
a number of connections of this to earlier analytical considerations.
%

First, we
should note that our approach here is based on
a particle picture that is most suitable to use in the semi-classical
or Thomas-Fermi limit of large $\mu$. As discussed, e.g., in Ref.~\cite{coles},
this is the regime where it is relevant to consider the vortex
as a ``particle'' without internal structure, characterized solely
by $(x_0,y_0)$.
On the other hand, earlier works
\cite{stoof,ashtone2}
have explored both dissipative and stochastic effects on the motion of
the vortex, but in a different regime wherein the wavefunction can be
approximated in a Gaussian form \cite{note}.
The latter is applicable closer to
the linear regime of the system. In light of these differences,
we will not attempt a direct comparison of these predictions, but we do
note the close proximity of the final result  of
Eqs.~(\ref{x_dot_revise})-(\ref{y_dot_revise}) with, e.g., Eq.~(25)
in Ref.~\cite{stoof}. Moreover, using the dimensional estimates presented
therein, we evaluate typical values of the parameter $\gamma$
in the regime $0.00023$--$0.0023$ for temperatures of the order
of $10$--$100$~nK.
Our present work will focus on the two cases of $\gamma = 0$ (no damping),
and $\gamma = 0.0023$, which represents an upper ``realistic'' limit for experiments,
beyond which the use of the DGPE may not be well justified for the particular physical system.
%

An additional class of techniques for deriving such effective
equations, also based on conservation laws, has been developed
from a rigorous perspective in Refs.~\cite{spirn,miot}.
However, the latter work considers settings where the vortices evolve on a homogeneous
background in the absence of a trap. More recently, these rigorous
methods have been explored in the presence of a trap as well
\cite{jerrard}.
In Appendix~\ref{appA} we provide an outline of how to utilize this class of methods in the
current setting. This, in turn, leads to a result similar to the one
obtained above in Eqs.~(\ref{x_dot_revise})-(\ref{y_dot_revise})
and thus further corroborates our theoretical predictions.

\subsection{Vortex Dipole Dynamics}

The above considerations can be generalized to the case of the vortex
dipole following a similar approximation as the one used in Ref.~\cite{pra_11}.
In particular, if we assume that the effect of interaction of the
``point vortices'' (in
the large chemical potential
regime) is independent of the anti-damped motion of each single vortex,
then the equations of motion combining the
two effects for the vortices constituting the dipole state read:
\begin{eqnarray}
\dot{x}_1 &=& -S_1\omega_{\rm pr,1}y_1-BS_2\frac{y_1-y_2}{2\rho_{12}^2}+s\gamma\dot{y}_1,
\label{center_x1}
\\
\dot{y}_1 &=& \phantom{-}S_1\omega_{\rm pr,1}x_1+BS_2\frac{x_1-x_2}{2\rho_{12}^2}-s\gamma\dot{x}_1,
\label{center_y1}
\\
\dot{x}_2 &=& -S_2\omega_{\rm pr,2}y_2-BS_1\frac{y_2-y_1}{2\rho_{12}^2}-s\gamma\dot{y}_2,
\label{center_x2}
\\
\dot{y}_2 &=& \phantom{-}S_2\omega_{\rm pr,2}x_2+BS_1\frac{x_2-x_1}{2\rho_{12}^2}+s\gamma\dot{x}_2,
\label{center_y2}
\end{eqnarray}
where the centers of the vortices are $(x_1,y_1)$ and $(x_2,y_2)$, with
respective charges $S_1=1$ and $S_2=-1$, and
$s$ defined similarly as in the case of the single vortex case.
For the rest of our analysis, in order to capture the precession
frequency increase as we depart from the BEC center and approach
its outer rim,
we will consider a slightly
modified form for the precession frequency for each
vortex used by Ref.~\cite{pra_11}:
\begin{equation}
\omega_{\rm pr,i}=\frac{\omega_{\rm pr}}{1-\frac{r_i^2}{R_{\rm TF}^2}}
\label{fett_eq},
\end{equation}
where $r_i=\sqrt{x_i^2+y_i^2}$ is the distance of each vortex
to the trap center, 
$\omega_{\rm pr}=\frac{\Omega^2}{2\mu} \log{\frac{A\mu}{\Omega}}$
is the precession frequency close to
the trap center, and $A=8.88$.
Finally, the coefficient
$B$ is a factor that takes into account the screening of the vortex
interaction due to the modulation of the density within the
Thomas-Fermi background cloud in which the vortices are seeded.
In the case of a homogeneous background, $B=2$
which will be adopted for cases of sufficiently large
chemical potential (such as $\mu=4$) herein.
For modulated
densities this value needs to be suitably modified; see,
e.g., Ref.~\cite{busch} for the form of the interaction in the latter case.
Here, adopting an approach similar to that of Ref.~\cite{middel2},
we
use an effective $B=1.35$ for computations with considerably lower chemical
potentials (such as $\mu \leq 1.6$).
This approach has been shown to yield accurate results for multi-vortices
in the case of $\gamma=0$~\cite{pra_11}.

The effective equations of motion for two opposite charge vortices
(\ref{center_x1})--(\ref{center_y2}) admit a steady state solution
corresponding to a stationary VD \cite{pra_11}.
The equilibrium positions for the VD are
$(x_{\rm eq},y_{\rm eq})=(0, \pm\frac{B}{4\omega_{\rm pr}+{B}/{R_{\rm TF}^2}})$.
In the next section we investigate the dynamics of single vortices and
VDs in the presence of the phenomenological dissipation and we
compare them with predictions from the analytical results
obtained in this section.

\begin{figure*}[htb]
\includegraphics[width=8.8cm]{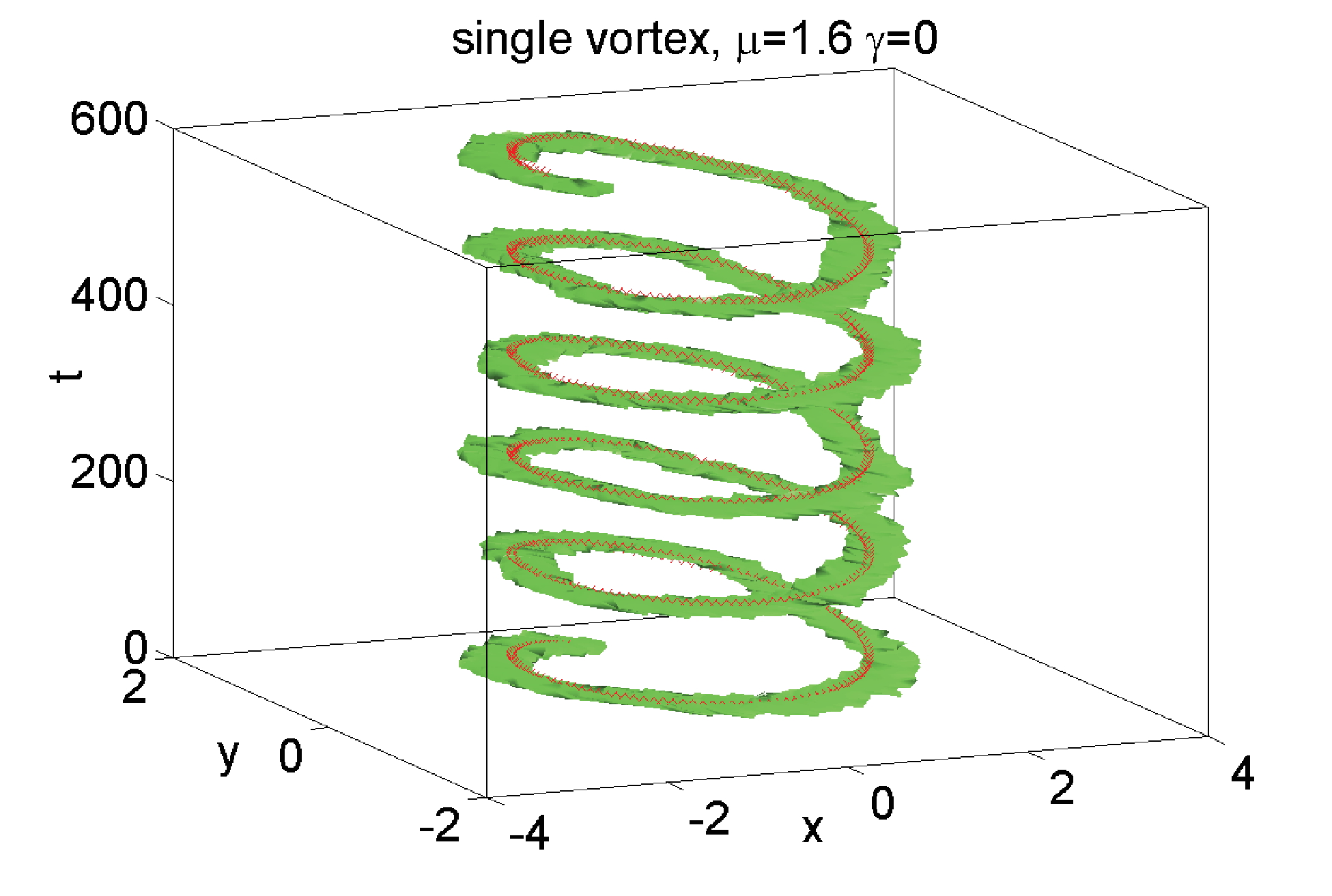}
\includegraphics[width=8.8cm]{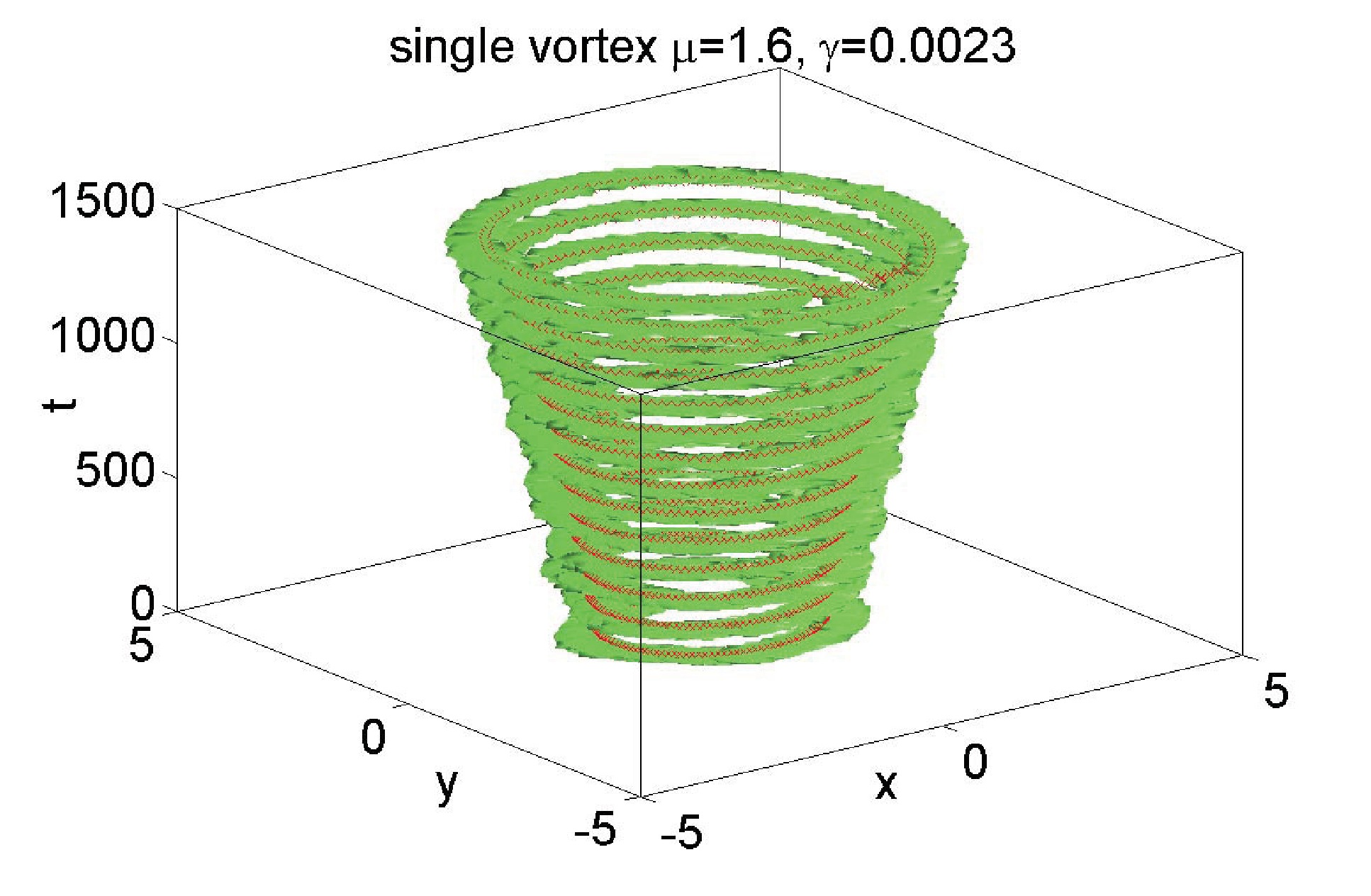}
\caption{
(Color online)
Isosurfaces of anti-damped single vortex motion
for $\gamma=0$ (left) and $\gamma=0.0023$ (right), showing
the full PDE [Eq.~(\ref{dissGPE})] numerical results (green)
and
the analytical result obtained from the vortex ODEs (red).
Here, the initial position is $(x_0, y_0)=(0, 1.5)$, the chemical potential
$\mu=1.6$ and the trap frequency $\Omega=0.2$.}
\label{density_dynamics_single_vortex_no_dissipation_mu_160}
\end{figure*}

We now turn to numerical computations to examine the validity of
our analytical approximations.

\section{Numerical Results}

We start by briefly discussing the key findings concerning the
predictions of the eigenfrequencies of the unstable system
and their dependence on the phenomenological dissipation parameter,
with further details left to Appendix~\ref{appB}.

Both in the cases of a single vortex and a symmetric vortex dipole, we observe good
agreement of our theoretical prediction with
the anomalous mode (complex) frequency associated with the vortex motion
over the regime of physical relevance of the DGPE; mathematically, this agreement actually
extends
far beyond the physically relevant $\gamma \ll 1$ regime,
a feature presumably associated with the non-perturbative
nature of our approach. On the other hand, we find the
method to be most accurate in the case of large chemical potential, where
the vortex can be characterized as a highly localized ``particle''
(a nearly point vortex without internal structure), while it is
less successful very close to the linear limit of the problem.
Moreover, the relevant complex pair of eigenfrequencies
corresponding to the anomalous mode {\it never} becomes real, i.e., the
motion always remains a spiral one independently of the strength of the
dissipation parameter. We highlight this here, as it is in contrast to what
is known in one-dimension for the case
of the dark soliton: in the latter
case, for sufficiently large $\gamma$, an over-damped
regime of exponential expulsion of the dark soliton
emerges~\cite{us}.

We now turn to the dynamical evolution of both the single vortex
and the counter-rotating vortex pair.
To that effect, we resort to direct numerical integration of
Eq.~(\ref{dissGPE}) for these states. In order to determine
the position of the vortex as a function of time, we compute the
fluid velocity
\begin{eqnarray}
v_s &=& -\frac{i}{2}\frac{u^*\nabla{u}-u\nabla{u^*}}{|u|^2}
\label{velocity}
\end{eqnarray}
and the fluid vorticity is defined as $\omega_{\rm vor}=\nabla\times v_s$. Since
the direction
of the fluid vorticity is always the $z$-direction, we can treat it as a
scalar. We can
determine the position of the vortex via a local center of mass of the
vorticity $\omega_{\rm vor}$ (around its maxima or minima). However,
in what follows,
we represent the vorticity iso-contours, which also enables us to explore
the full (2+1)-dimensional space-time motion $(x,y,t)$ of the vortex.

Figure~\ref{density_dynamics_single_vortex_no_dissipation_mu_160}
compares the
full space-time vorticity dynamics of the single vortex for
$\gamma=0$ (i.e., the Hamiltonian case of pure condensate) with the
``finite temperature'' case
of $\gamma=0.0023$.
The figure compares the actual vorticity isocontours (thick green lines)
to the theoretical prediction based on our ODEs
of Eqs.~(\ref{x_dot_revise})-(\ref{y_dot_revise}) for the respective
$\gamma$ (red lines). On the whole, agreement for the single vortex case is
extremely good in all cases studied. As evident from the figure
for any $\gamma>0$, the ``particle'' ODEs yield a
well-defined anti-damped pattern, whereas the full numerical simulation reveals a small
amount of modulation on top of the underlying trajectory ---this is presumably related to
the (nonlinear) role of continuous vortex-sound interactions within a trap (anticipated  to be at most
of order few $\%$ \cite{ft1}) which is not accounted for in our analytical model.
In addition to this, closer inspection reveals a very slight deviation
(typically less than $1\%$ in the case of a single vortex) in the evolution timescales, associated
with tiny shifts in both the amplitude and the frequency of the anti-damped motion.
As anticipated, the agreement improves further with increasing values of $\mu$, and as such deviations are not expected to
lead to any experimentally noticeable effects, they will not be discussed any further here.

\begin{figure*}[htb]
\begin{tabular}{cc}

\includegraphics[width=9cm]{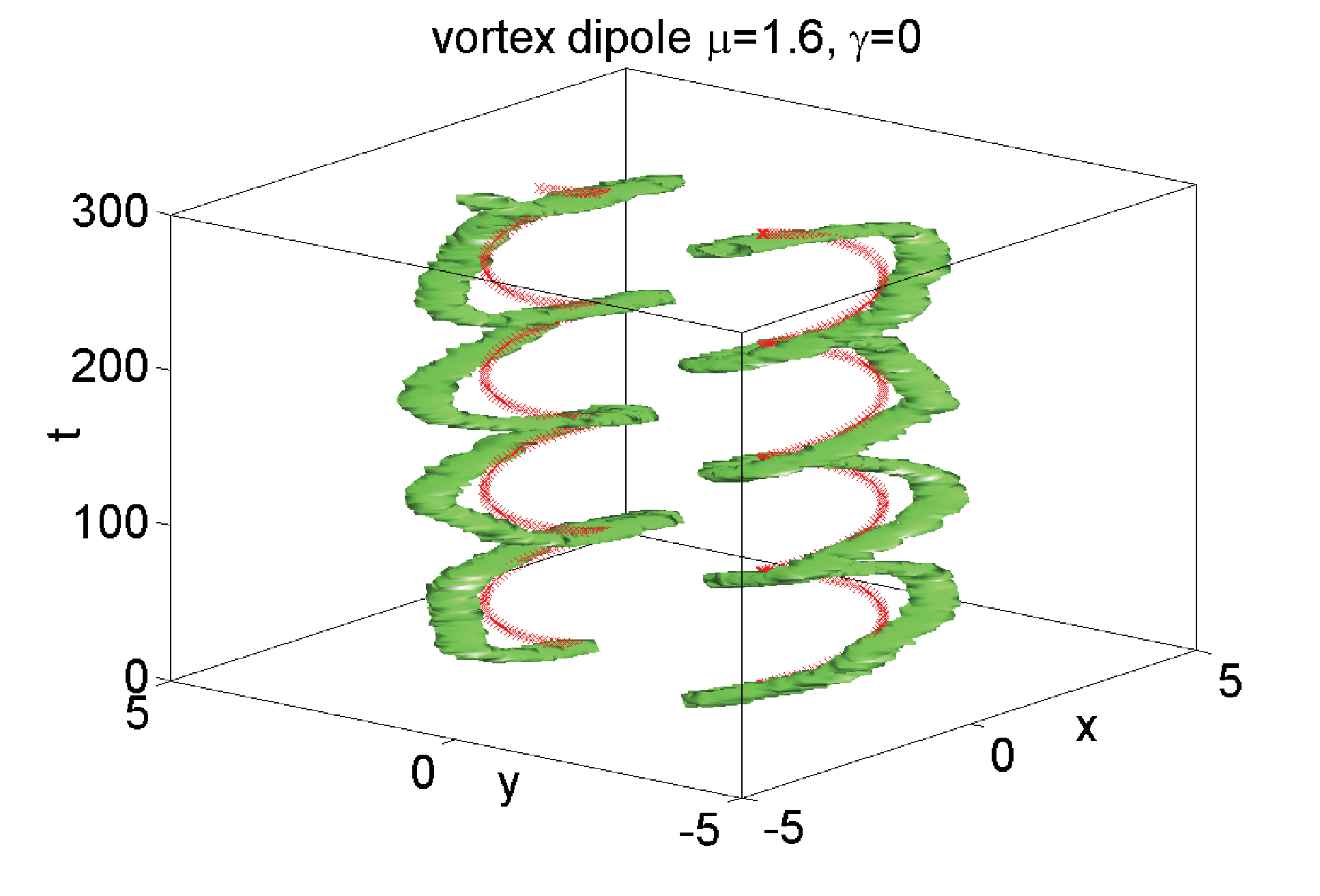}
\includegraphics[width=9cm]{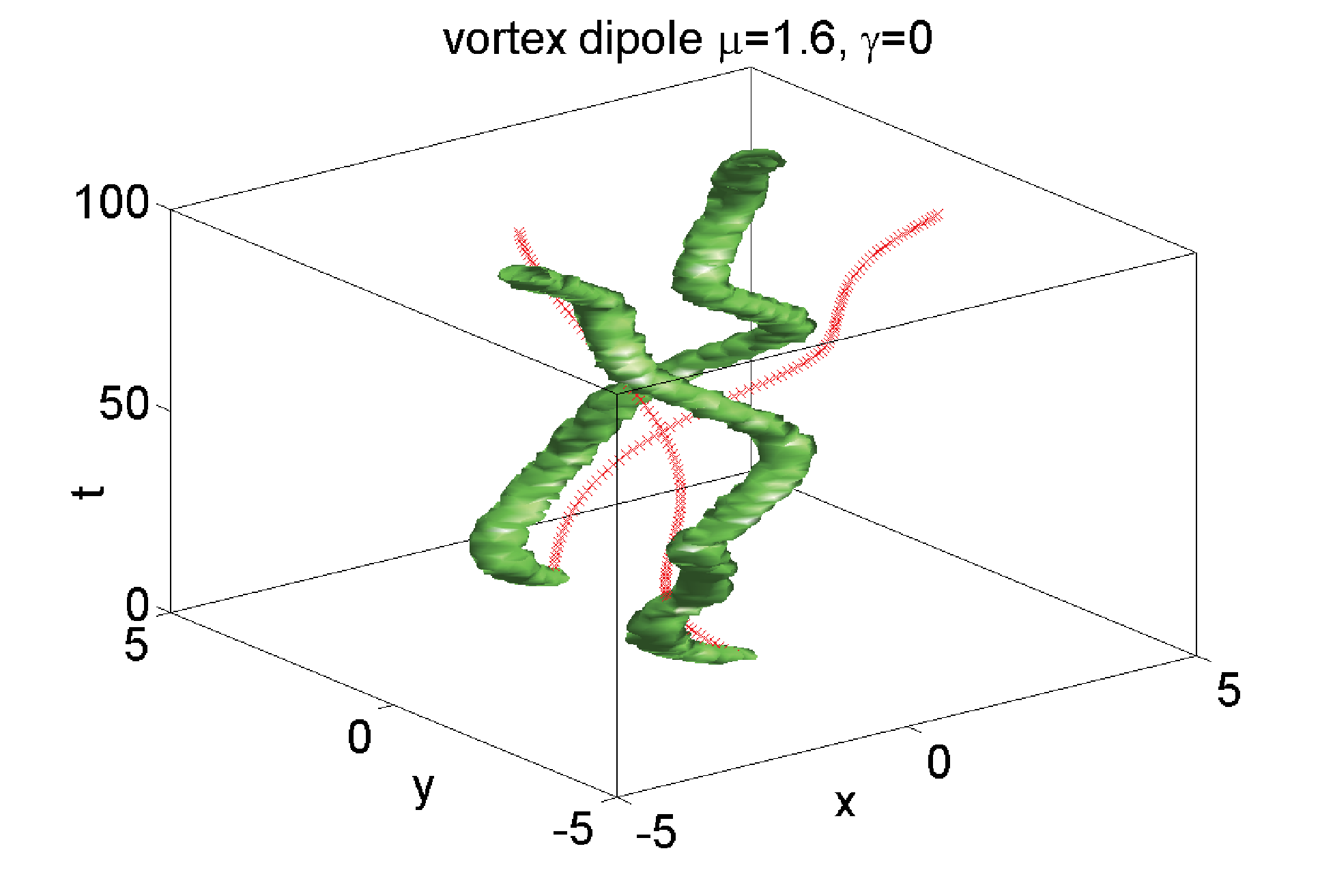}
\\
\includegraphics[width=9cm]{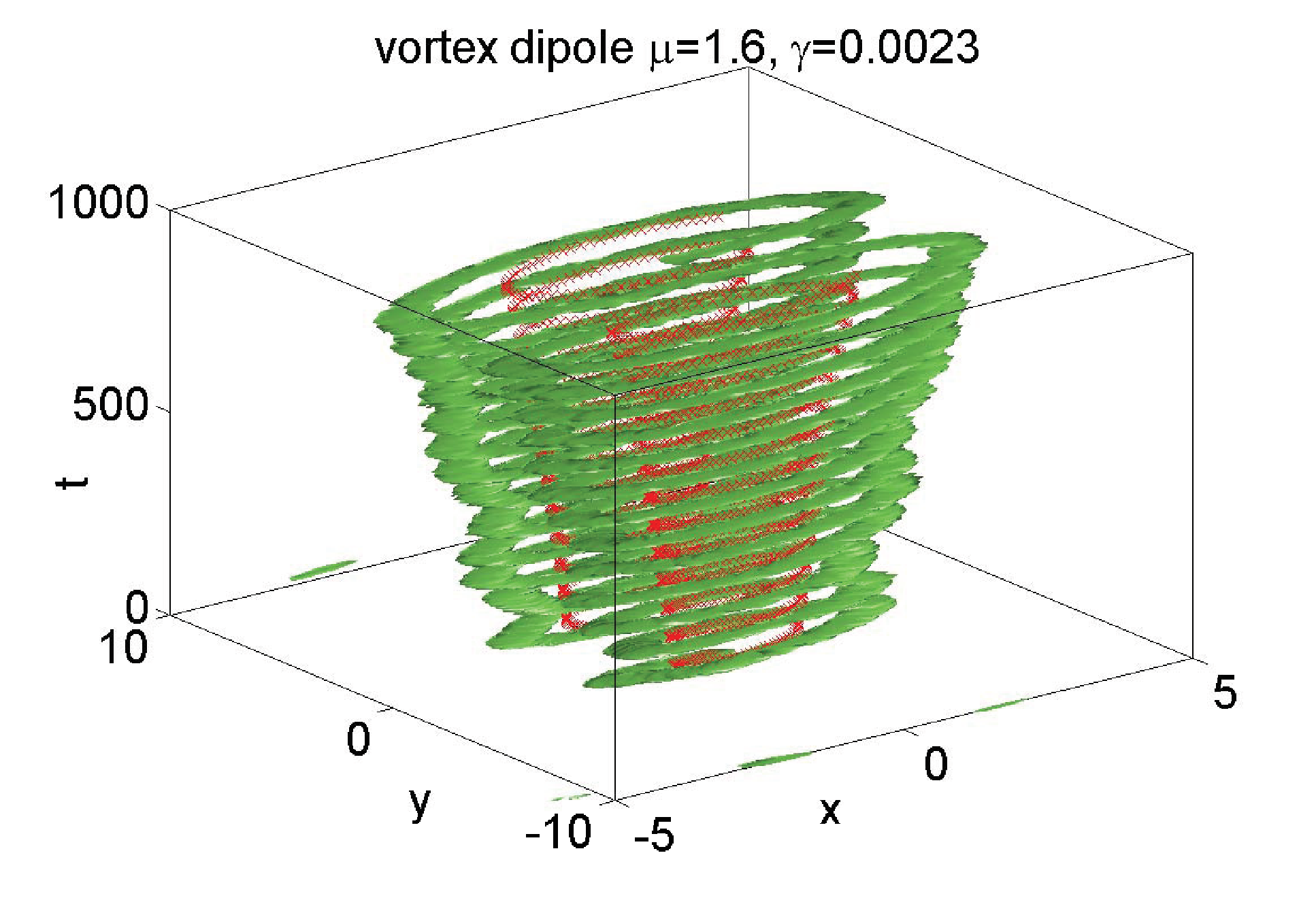}
\includegraphics[width=9cm]{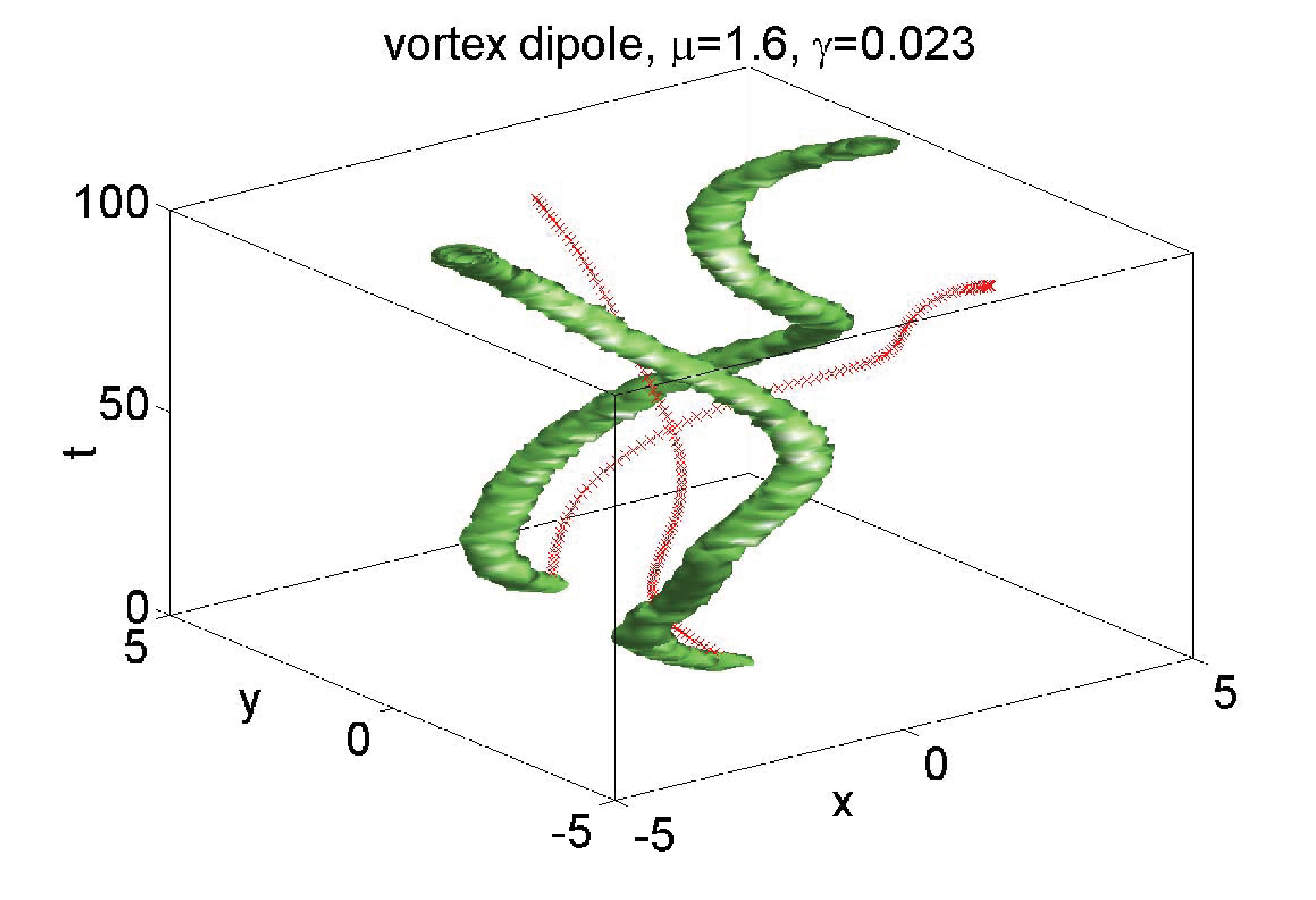}
\end{tabular}
\caption{
(Color online)
Motion for a symmetric (left column) and an asymmetric (right column)
vortex dipole in the absence (top panels) and presence (bottom panels)
of dissipation.
The
initial positions of the two vortices are:
Left Column: $(x_1, y_1)=(0, 1.75), (x_2, y_2)=(0, -1.75)$,
Right Column: $(x_1, y_1)=(0, 3.00), (x_2, y_2)=(0, -1.25)$,
while the chemical potential
$\mu=1.6$ and the trap frequency $\Omega=0.2$. Other parameters as in
Fig.~\ref{density_dynamics_single_vortex_no_dissipation_mu_160}.}
\label{density_dynamics_vortex_dipole_epicycle_no_dissipation_mu_160}
\end{figure*}

The situation becomes slightly more complicated in the case of a vortex dipole
(Fig.~\ref{density_dynamics_vortex_dipole_epicycle_no_dissipation_mu_160}), as both
the vortex anti-damped motion and the vortex-sound interactions acquire an additional channel associated
with the vortex-vortex (and, indirectly, sound-sound) interactions \cite{prouk12}.
Here we need to distinguish two cases, based on the symmetry of the case under study (or its absence).

\begin{figure*}[htb]
\begin{tabular}{cc}
\includegraphics[width=8.5cm]{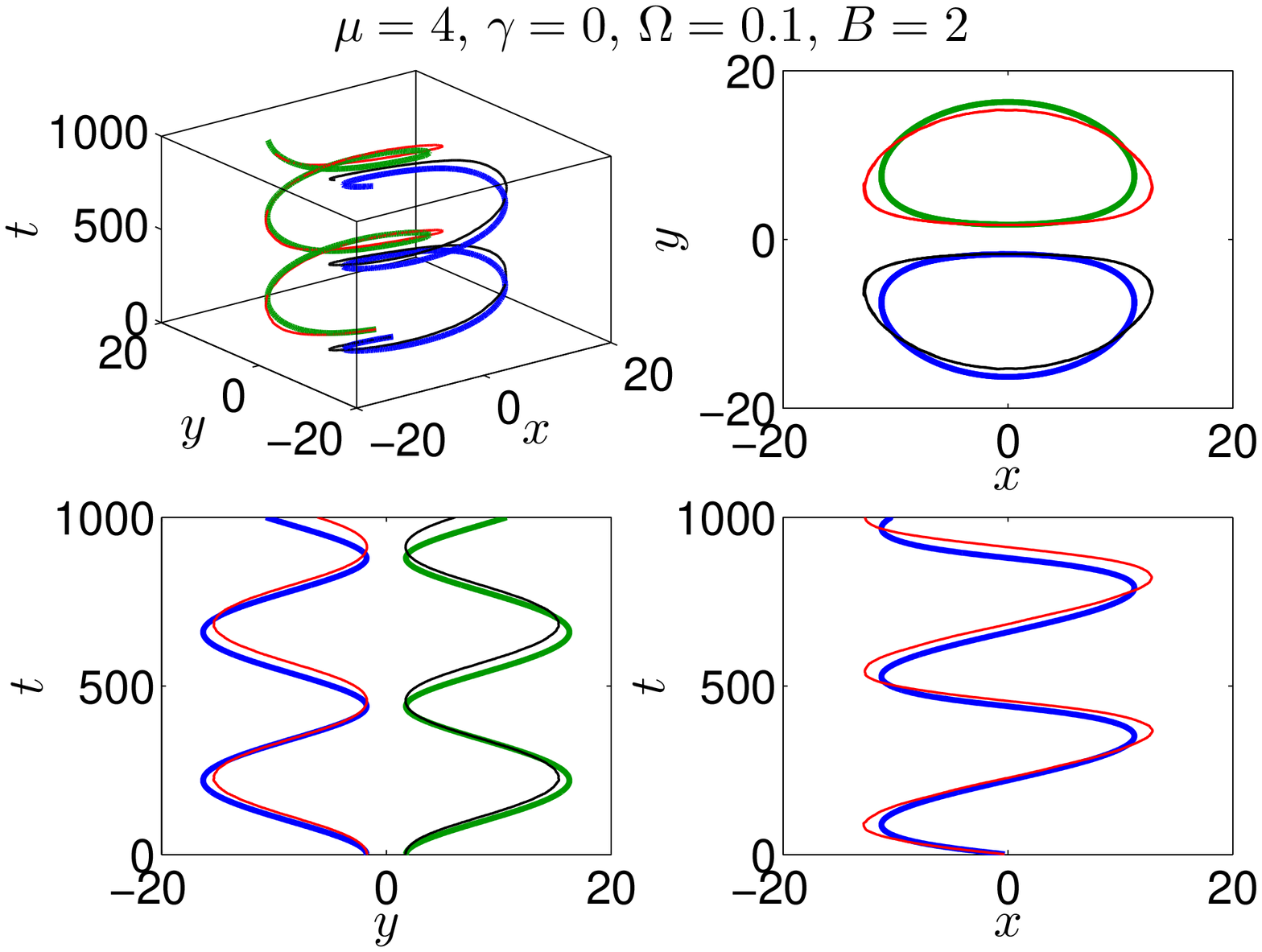} ~&~
\includegraphics[width=8.5cm]{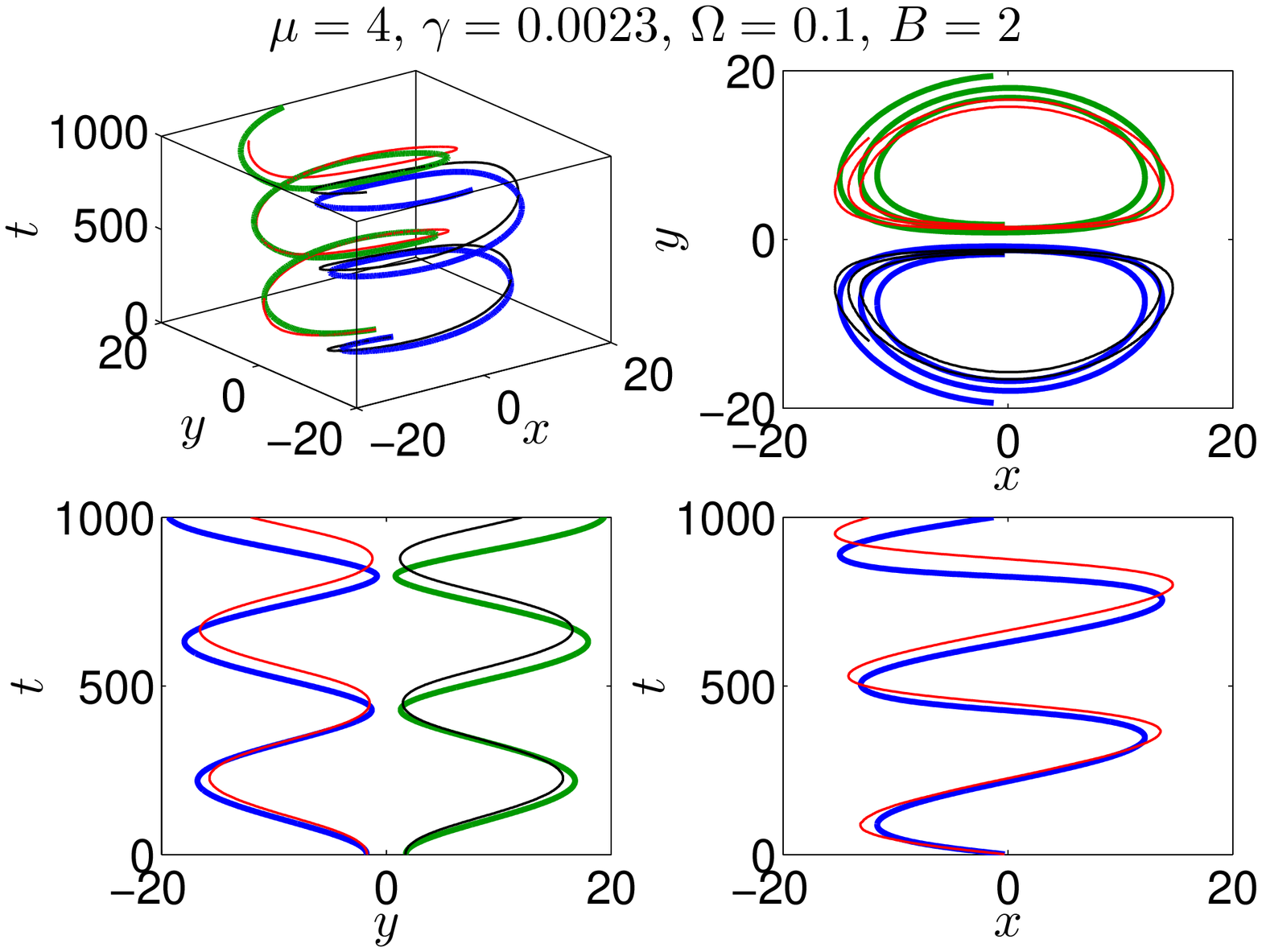}
\end{tabular}
\caption{
(Color online)
Left four panels: Analytical versus numerical trajectories of the vortex dipole epicycle
for $\mu=4$, $\gamma=0$, $\Omega=0.1$, $B=2$ case.
In particular, the upper left subpanel is the three-dimensional comparison of the vortex instantaneous positions
$(x,y)$ versus $t$,
the upper right subpanel is the
projection of the motion on the $(x,y)$ plane, the
bottom left subpanel is the projection on the $(y,t)$ plane and the bottom right
subpanel is the projection on the $(x,t)$ plane.
Here the thick blue and green lines correspond to the analytical predictions
while the thin red and black lines correspond to the numerical results.
Right four panels: As in the left panels, but with
$\gamma=0.0023$.}
\label{mu_4_omega_01_gamma_0_epicycle_b_2_adjusted2}
\end{figure*}

Initially, we consider the case of a {\it symmetric} vortex dipole
(Fig.~\ref{density_dynamics_vortex_dipole_epicycle_no_dissipation_mu_160}, left column), focusing for consistency on
the same parameters as for the single-vortex case
(Fig.~\ref{density_dynamics_single_vortex_no_dissipation_mu_160}), where excellent agreement has already been reported;
here, we effectively excite the
epicyclic precession mode of the vortex dipole.
We find that for the relatively small chemical potentials $\mu \sim O(1)$ considered here, there is a small noticeable deviation even for $\gamma=0$; this is presumably due to the neglect of vortex-sound interactions in the analytical model.
We note that in the case of finite $\gamma$,
the epicyclic trajectory will continue to expand outward until the vortices
essentially merge with the vanishing background of the cloud
and disappear thereafter;
%
here, the corresponding analytical anti-damped pattern ($\gamma \neq 0$) consistently predicts a slightly lower amplitude of oscillations, with the discrepancy increasing with increasing $\gamma$ (at constant small value of $\mu$). However, a slight increase in the value of $\mu$  can considerably suppress such differences (see subsequent Fig.~\ref{mu_4_omega_01_gamma_0_epicycle_b_2_adjusted2} and related discussion).


Fig.~\ref{density_dynamics_vortex_dipole_epicycle_no_dissipation_mu_160} also shows the case of an {\it asymmetric} vortex dipole (right columns), revealing that this
deviation becomes more pronounced in this case, with an increase in the relative difference in off-centered locations of the two vortices amplifying such an effect. To display this effect, we have chosen to show here a highly asymmetric initial VD configuration, which nonetheless yields reasonable qualitative agreement.

In particular, Fig.~\ref{density_dynamics_vortex_dipole_epicycle_no_dissipation_mu_160} (top right panel)
shows that in the
Hamiltonian case of $\gamma=0$, one of the vortices rotates closer
to the trap center, while the other one precesses further outside.
On the other hand, in the presence of anti-damping
(bottom right panel), the vortices
rapidly spiral towards the Thomas-Fermi radius and cannot be accurately
tracked thereafter. It is important to note that in this case, involving
dynamics of the vortices fairly close to the Thomas-Fermi radius,
we have generally found
(in dynamical evolutions not shown
herein) that our theoretical analysis and particle model are least likely
to properly capture the vortex dynamics. This is because in this case
the condensate's density modulation in fact affects most significantly
both of the dynamical elements contained in our model, namely the precession
and the inter-vortex interaction. As regards the precession, we have found that
Eq.~(\ref{fett_eq}) is progressively less accurate for distances
approximately satisfying $r > 0.7 R_{\rm TF}$. At the same time, this
density modulation most significantly affects the screening
effect discussed above (and detailed, e.g., in Ref.~\cite{busch}).
As a result, in settings such as those of
Fig.~\ref{density_dynamics_vortex_dipole_epicycle_no_dissipation_mu_160} (right column)
we may expect a rough qualitative agreement between the numerical observations
and our simple ODE model, but a quantitative match would require
a considerably more complex particle model accounting for the above
traits, which 
is beyond the scope of the present work.

We note that this trend seen in the dependence of the VD motion is in qualitative agreement
with previous findings in the related case of two interacting dark solitons in a single harmonic trap. In particular,
Ref.~\cite{allen_soliton}, highlighted the increased role of soliton-sound interactions for two solitons located
in {\it asymmetric} positions in a one-dimensional harmonic trap, which allowed significant energy exchange.


In the spirit of exploring the relevance of the model to other parametric
regimes, we have also attempted to probe similar features, as e.g.~in
Fig.~\ref{density_dynamics_vortex_dipole_epicycle_no_dissipation_mu_160},
for different sets of parameter values. Our conclusion from
this study is that for larger values of the chemical potential
and smaller values of the trap frequency $\Omega$, the model becomes
generally and progressively more accurate. This is confirmed not
only by the eigenvalue computations and more accurate linearization
predictions (see Appendix~\ref{appB}), but also by direct numerical computations
of the vortex dynamics as, e.g., in the dipole epicyclic evolution
of Fig.~\ref{mu_4_omega_01_gamma_0_epicycle_b_2_adjusted2}. The figure
shows two epicyclic evolution examples for the cases of $\gamma=0$
(left) and $\gamma=0.0023$ (right), while reducing $\Omega$ to $0.1$
and increasing the chemical potential $\mu$ to $4$~\cite{footnote2}.
Both the fully three-dimensional
evolution of $(x,y)$ as a function of $t$ and the cross sections in
the $(x,y)$, $(x,t)$ and $(y,t)$ planes reveal a very accurate capturing
of the motion of the vortices. Nevertheless, we should note here
that even slight ``detunings'' from the exact vortex motion (e.g.
due to the imperfect capturing of the screening effect) lead to
a cumulative effect over the long integration times displayed
e.g. in Fig.~\ref{mu_4_omega_01_gamma_0_epicycle_b_2_adjusted2}.
This renders more difficult (and, arguably, less meaningful) the
introduction of a quantitative measure of the proximity of the
trajectories as measured at any given time instance, such as a relative
position error.

\section{Conclusions and Future Challenges}

In this work, we have provided an energy-based, semi-analytic
method for deriving the effective dynamics of vortices in the presence
of both 
an inhomogeneous background, and importantly, a damping term that accounts
phenomenologically for the qualitative effect of the thermal cloud.
In the context of the simple, yet accurate at sufficiently low temperatures,
dissipative Gross-Pitaevskii equation,
the single vortices were confirmed (in accordance with
earlier reported works) both analytically and numerically to
spiral outwards towards the rim of the
condensate, disappearing in the corresponding background.
While this DGPE model is only valid far from the critical temperature due to
the result of ignoring fluctuations, this is nonetheless a relevant
regime for numerous experiments given the
excellent experimental control currently available.
Starting from such a model (and ignoring  stochastic
fluctuations which are expected to become significant at higher temperatures
and closer to the critical region as discussed, e.g., in Refs.~\cite{stoof,ashtone2}),
our considerations provide an
explicit analytical description of
such spiraling (for a given dissipation parameter), enabling the quantification of the relevant effect in the Thomas-Fermi regime where the vortex can
be considered as a ``particle''.
In fact, one can envision a ``reverse'' process, whereby from experiments
such as the ones of Ref.~\cite{freilich}, one can quantitatively infer
an ``effective'' value of such a dissipation coefficient
$\gamma$, through comparison with the analytical
predictions of Eqs.~(\ref{x_dot_revise})-(\ref{y_dot_revise}).
Importantly,
we have subsequently generalized such considerations
to the case of a vortex dipole, a setting of intense
recent physical interest also experimentally~\cite{BPA_10,pra_11}.
We have illustrated in the latter how the internal epicyclic
motion of the vortices is also converted into an outward spiraling.
In this setting, the analytical approximation provided
a qualitative description of the corresponding dynamics,
with deviations becoming generally less pronounced with larger condensates
and more symmetric initial vortex configurations.

The present work paves the way for a multitude of future
possibilities. On the one hand, one can consider more complex
two-dimensional settings including larger vortex clusters,
and co-rotating vortex patterns (see recent experimental and theoretical results in Ref.~\cite{corot}),
instead of purely counter-rotating
ones, such as the dipoles considered herein.
An additional direction that could be very relevant to explore,
particularly in the case of asymmetric vortex dipoles,
concerns the interplay of vortices with sound waves
as the sound emitted from each of the
vortices could be affecting the motion of the other vortex within the
pair, in a way reminiscent of the recent analysis of Ref.~\cite{prouk12}.
On the other hand,
a very natural extension would be to attempt to provide effective
equations for a single vortex ring in the context of a three-dimensional space-time
generalization of the dissipation considered herein. Such vortex
rings may appear even spontaneously in three-dimensional space-time
BECs~\cite{bpa_vr} and even their
interactions have been experimentally monitored~\cite{hau}
(see also the relevant review of Ref.~\cite{komineas}), hence
the development of a particle based formulation for their motion
seems like a particularly exciting topic for future study.

\acknowledgments{
We gratefully acknowledge the support of NSF-DMS-0806762
and NSF-DMS-1312856, NSF-CMMI-1000337, as well as from
the AFOSR under grant FA950-12-1-0332, the
Binational Science Foundation under grant 2010239, from the
Alexander von Humboldt Foundation and the FP7, Marie
Curie Actions, People, International Research Staff
Exchange Scheme (IRSES-606096).
}

\appendix

\section{An Alternative approach to the vortex center dynamics}
\label{appA}

We start by considering the PDE of the form of Eq. \eqref{dissGPE}.
We then define the following quantities
\begin{align}
j(u) & :=  - {i \over 2} \LC u^* \nabla u - u \nabla u^* \RC ,  \\
J(u) & := {1 \over 2} \curl j(u) .
\end{align}
which correspond, respectively, to the
momentum density and (half of) the vorticity, but we will treat
them here as mathematical quantities.

\begin{figure*}[hbt]
\includegraphics[height=5cm]{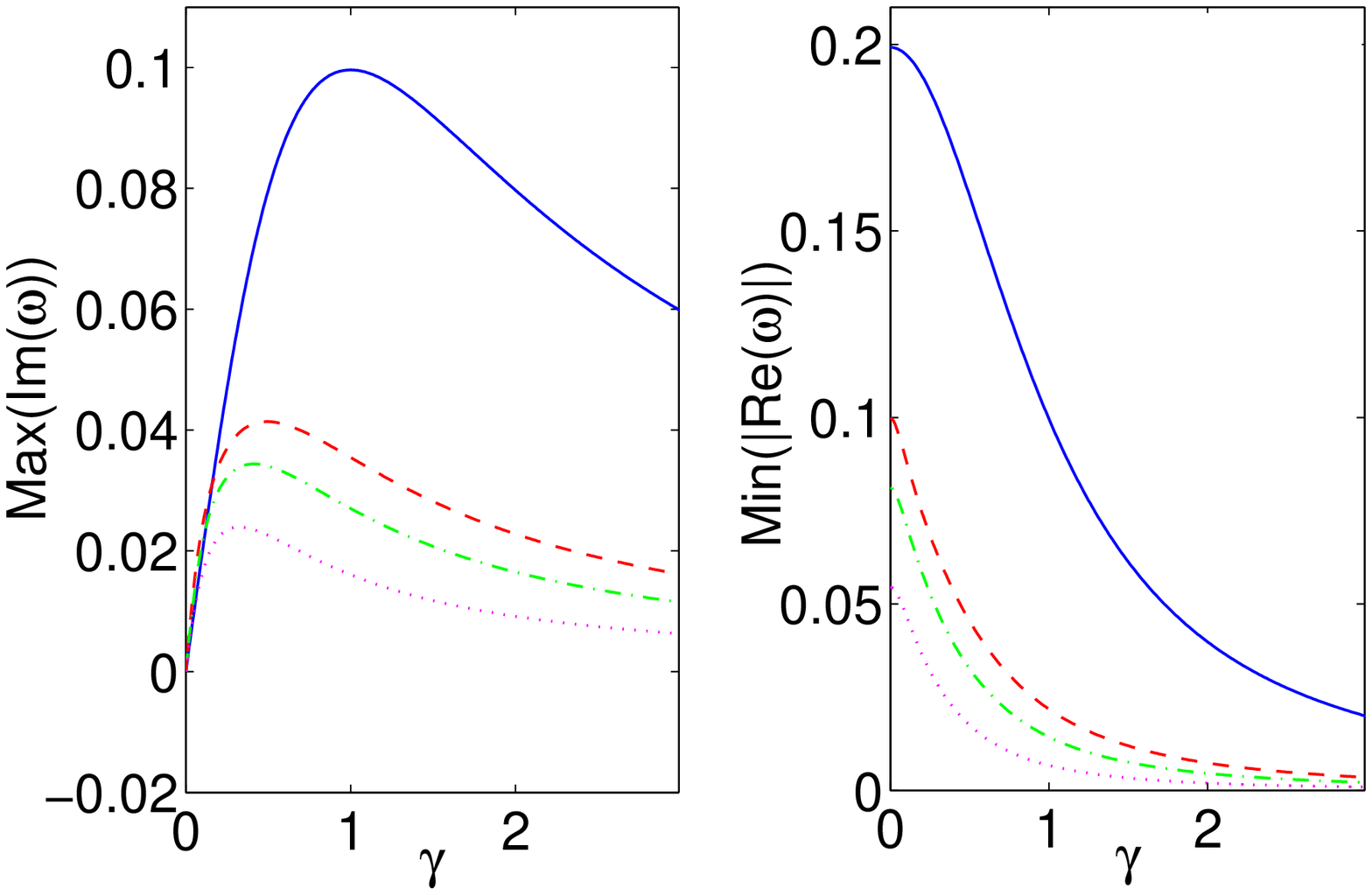}
~~~~~~~~~~~
\includegraphics[height=5cm]{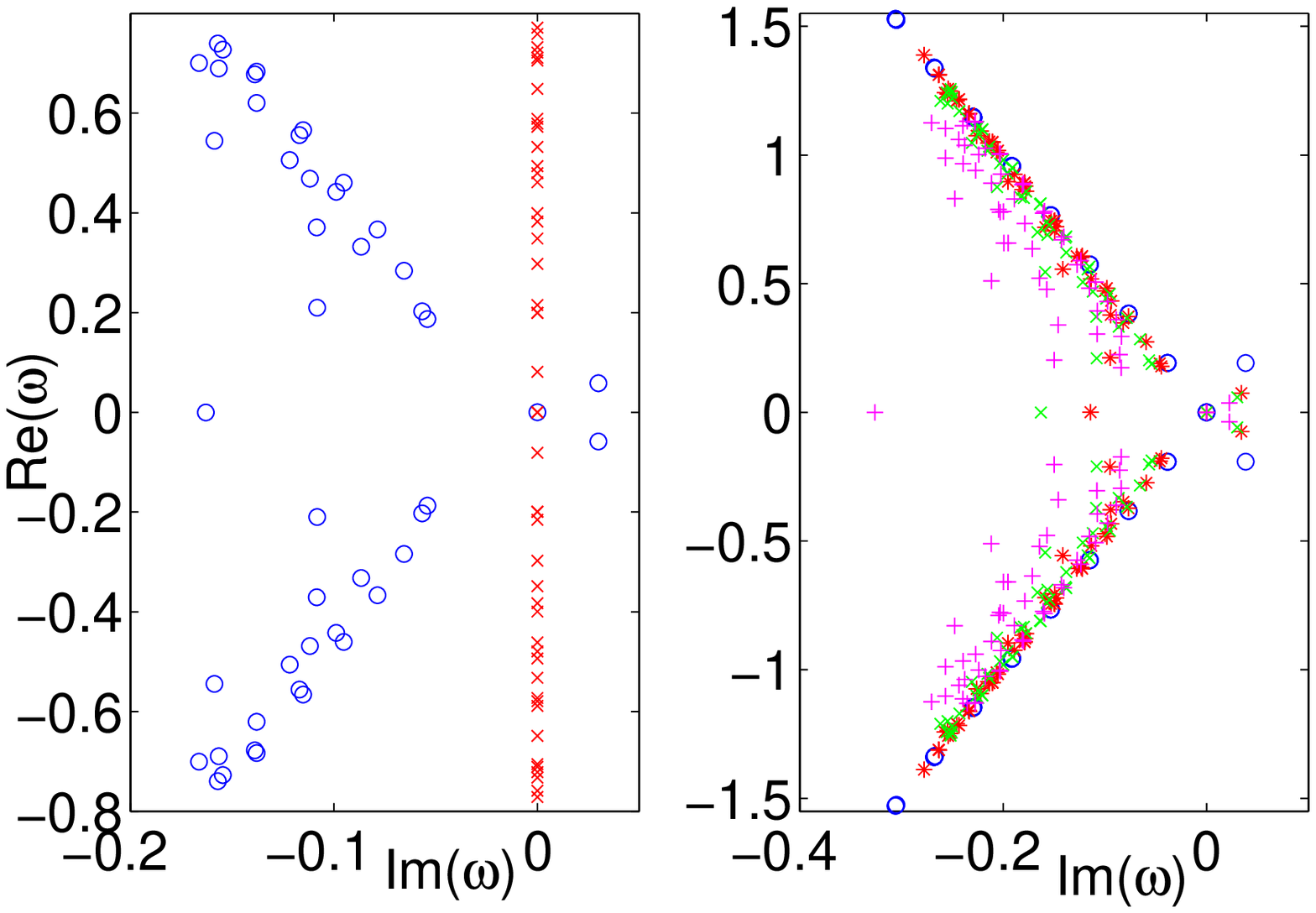}
\caption{
(Color online)
Left Images: The left subpanel shows the non-vanishing imaginary part of the eigenfrequency
associated with the anomalous mode, as soon as the dissipative
prefactor $\gamma$
is non-zero for $\mu=0.4$, $0.8$, $1$, and $1.6$ (respective
curves from top to bottom).
The right subpanel is similar but now for
the real part of the eigenfrequency. Here $\Omega=0.2$.
Right Images: The left subpanel illustrates the linearization (so-called
BdG) spectrum of the case $\gamma=0$ (red crosses)
and that of $\gamma=0.2$ (blue circles) for $\mu=1$ and $\Omega=0.2$.
The right subpanel
shows the spectrum for $\mu=0.4$ (blue circles), $\mu=0.8$ (red stars), $\mu=1$
(green crosses) and $\mu=1.6$ (magenta pluses) for $\gamma=0.2$.
}
\label{dissipation_mu_vary_compare}
\end{figure*}

It can then be shown directly  that:
\begin{equation} \label{Jacobianequation}
\begin{split}
\p_t J(u) & = \gamma \curl (\p_t u, \nabla u) -{1\over 2}  \curl \dv ( \nabla u \otimes \nabla u ) \\
& \quad  - {1\over 2}
\curl (|u|^2  \nabla V),
\end{split}
\end{equation}
where $(a, b) ={1\over 2}(  a b^* + a^* b)$ is the complex inner-product and $\otimes$ denotes the tensor product.

We will now examine the magnitude of the different terms in
the identity~(\ref{Jacobianequation}) by direct calculation
based on an ansatz of the form $u(x,t) = \sqrt{\rho_{\rm TF}(x)} v(x-a(t))$
where $v=\sqrt{\rho} e^{i \theta}$ will denote the vortex of the
homogeneous problem; in turn, the Thomas-Fermi density
{$\rho_{\rm TF}(x) = \mu( 1 - {|x|^2 \over R^2_{\rm TF}})$}.
Then we have for the {momentum} term:
$j(v)(x) = \rho(x) \nabla \theta(x) = {1\over |x|} \rho(|x|) \tau$ where
$\tau = (-y,x)/|x|$.
Notice that for the spatial scales of interest to this calculation,
we will have:  $r = (2 \mu)^{-{1\over 2}}$ with $r, |a| \ll R_{\rm TF}$, i.e.,
$r$ denotes a spatial scale of the order of the healing length of
the vortex. In particular $\rho(r) \approx 1$. The symbol $a$ will be used to denote the vector
position of the vortex, which is assumed to be well within the
Thomas-Fermi radius $R_{\rm TF}$.

Integrating $J(u)$ against a spatial variable $x_k$ (where the subscript index
$k=1, 2$ denotes the two dimensions) yields
\begin{align*}
\int_{B_r(a)} x_k J(u) dx
& = {\rho^2(r) \over 2 r} \int_{\p B_r } (x_k + a_k) \rho_{\rm TF}(x + a) d \ell\\
& \quad +  {1 \over 2} \int_{B_r} \rho_{\rm TF}(x+ a) \rho(|x|) \nabla \theta \times e_k dx.
\end{align*}
Here $B_r(a)$ is a ball of radius $r$ around the center position $(a)$
of the vortex and $\p B_r(a)$ denotes the boundary of such a region.
The first among these terms yields:
\begin{eqnarray*}
{\rho(r)  \mu \over 2 r} \LB  2 \pi r a_k  \LB  1 - {r^2 +  |a|^2 \over R_{\rm TF}^2} \RB  -{\pi r^3 a_k  \over R_{\rm TF}^2}   \RB  \\
= \pi  \rho(r) \mu a_k \LC 1 - {|a|^2 \over R_{\rm TF}^2} \RC + O\left( {\mu r^2 |a| \over R_{\rm TF}^2}\right),
\end{eqnarray*}
and the second term  is found to be also of $O({ \mu r^2 |a| \over R_{\rm TF}^2})$.
From the time derivative of the Jacobian term, we thus obtain to leading
order:
\begin{eqnarray}
\partial_t \int_{B_r(a)} x_k J(u) dx = \pi \rho(r) \mu \dot{a} .
\label{extra1}
\end{eqnarray}

A longer calculation for the moment of the dissipative term in
Eq.~(\ref{Jacobianequation}) leads
to an integral of the form $\int_{B_r(a)} x_k \curl ( \partial_t u,
\nabla u ) dx$.  Based on the ansatz, we get a dominant contribution
of the form
\begin{equation} \label{}
 - \gamma  \mu \LB  {1\over 2 } \int_{B_r} \LV \nabla v \RV^2 dx - \pi \rho(r) \RB \dot{a} \times e_k  
 \end{equation}
 with error terms bounded by {$O({\mu |a|^2 |\dot a| \over R_{\rm TF}^2})$}.

The interaction term can be reshaped using the following identity:
$\curl \dv ( \nabla u \otimes \nabla u) = \curl \LB (\Delta u, \nabla u) + {1\over 2} \nabla |\nabla u|^2 \RB = \curl ( \Delta u , \nabla u)$, and a calculation shows
\begin{align*}
 & \int_{B_r(a)} x_k \curl \dv \LC \nabla u \otimes \nabla u \RC dx \\
& \qquad = \int_{ \p B_r} (x_k + a_k)    \LC \Delta u , \p_\tau u \RC d \ell \\
& \qquad \quad + \int_{ \p B_r} \LC \p_\nu u , \nabla u \times e_k \RC d \ell \\
&\qquad  \quad  - \int_{ \p B_r}{1\over 2} \tau \cdot e_k \LV \nabla u \RV^2 d \ell .
\end{align*}
Using that $\int_{\p B_r} (\p_\nu v, \p_k v) d \ell = \int_{\p B_r} \tau \cdot e_k |\nabla v|^2 d \ell = 0$, then a similar calculation shows that  in the absence of a second vortex the terms on the right hand side can be bounded
by $O({\mu \over R_{\rm TF}^2})$.

Finally, a critical contribution is made by the potential term.
If $V(x) = {\Omega^2 \over 2} |x|^2$ then $\p_j V(x) = \Omega^2 x_j$ and
\begin{align*}
& - {1\over 2} \int_{B_r(a)}  x_k \curl (|u(x)|^2 \nabla V)   dx \\
& = - {\Omega^2 \over 2}  \int_{\p B_r }(x_k + a_k)  |u(x+a)|^2   \tau \cdot a   \, d \ell \\
& \quad -{ \Omega^2 \over 2}  \int_{ B_r }  |u(x+a)|^2    (x+a) \times e_k  \, d x
\end{align*}
which comes to
\begin{equation}\label{potentialest}
 {\Omega^2 \over 2} \mu \LB \pi \rho(r) r^2 -\int_{B_r} \rho(x) dx \RB (a\times e_k)
\end{equation}
up to an error of $O({\Omega^2 \mu  |a|^5 \over R_{\rm TF}^2})$.

Putting the different contributions together from Eqs \eqref{extra1}--\eqref{potentialest} and dividing by $\pi \rho(r) \mu = \pi \mu$, we arrive
at an effective ODE for the motion of the vortex,
\begin{align*}
&  \dot{a} + \gamma A (\dot{a} \times e_k )  = -  {\Omega^2 \over \mu}  B  {( a \times e_k) },
\end{align*}
where
\begin{align*}
A & = {1\over 2 \pi } \int_{B_{(2\mu)^{-{1\over2}}}} \LV \nabla v \RV^2 dx - 1 , \\
B & =  {1 \over 4 \pi \mu^{-1}}\int_{B_{(2\mu)^{-{1\over2}}}} \rho(x) dx - 1.
\end{align*}
Since $(1 - {|a|^2 \over R_{\rm TF}^2}) \dot{a} = \dot{a}$ up to $O({1\over R_{\rm TF}^2})$, we
achieve the
analogous ODE to the result of Eqs.~(\ref{x_dot_revise})-(\ref{y_dot_revise}),
up to terms logarithmic in $\mu/\Omega$ and errors of $O({1\over R_{\rm TF}^2})$.  Thus, this higher order moment method yields a similar conclusion to the
one obtained by the energy methods earlier in the text.
In order to discern the higher order corrections using this method, one should look at the situation when the healing length is significantly smaller than the vortex position, $r \ll |a|$.

\section{Stability analysis of vortices and vortex dipoles in the DGPE framework}
\label{appB}

We present here some further
details related to the eigenfrequency spectrum
of the single vortex and vortex dipole cases; for mathematical completeness,
we present our results for values of $\gamma$ in the range $0 \le \gamma \le 3$,
noting that values of
$\gamma$ well beyond the interval outlined previously (in the main text)
have no physical relevance for the particular problem.

First, Fig.~\ref{dissipation_mu_vary_compare} (left plots) depicts real and imaginary parts
of the eigenfrequency associated with the anomalous mode of
the vortex centered at the origin
for different values of $\mu$ versus $\gamma$
for the DGPE [Eq.~(\ref{dissGPE})].
This shows that the mode has a positive imaginary part of the
eigenfrequency (i.e., a real part of the corresponding eigenvalue, directly
signaling the relevant instability) for any
non-zero value of $\gamma$.
In Fig.~\ref{dissipation_mu_vary_compare} (right plots), we
show how the spectrum of the Hamiltonian case ($\gamma=0$) gets modified by
a $\gamma \neq 0$ term 
depicting also the ``trajectory'' of the relevant
eigenvalues of the spectrum 
for different values of the chemical potential $\mu$.

\begin{figure*}[hbt]
\includegraphics[width=8cm]{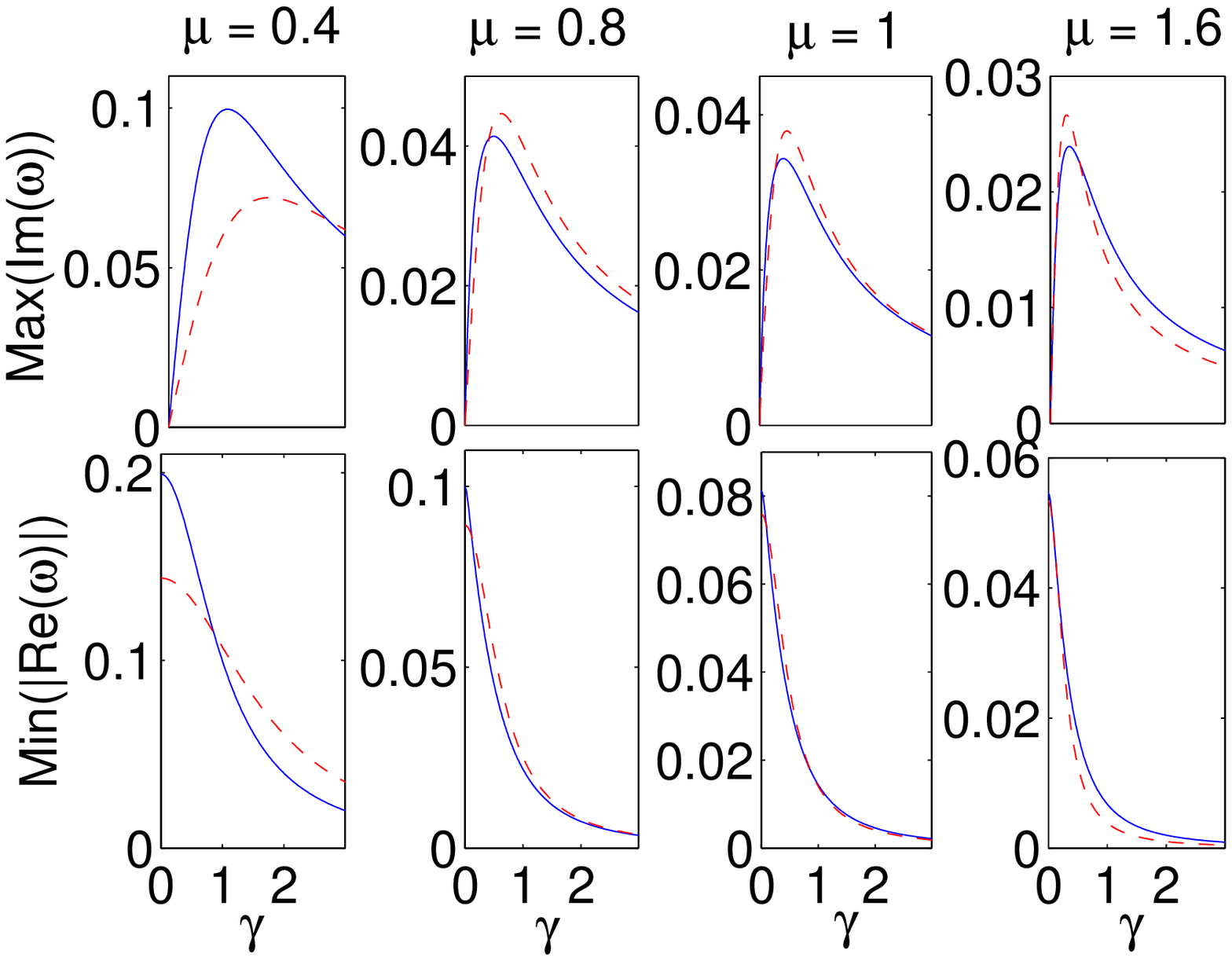}
~~~~~~~~~~~
\includegraphics[width=8cm]{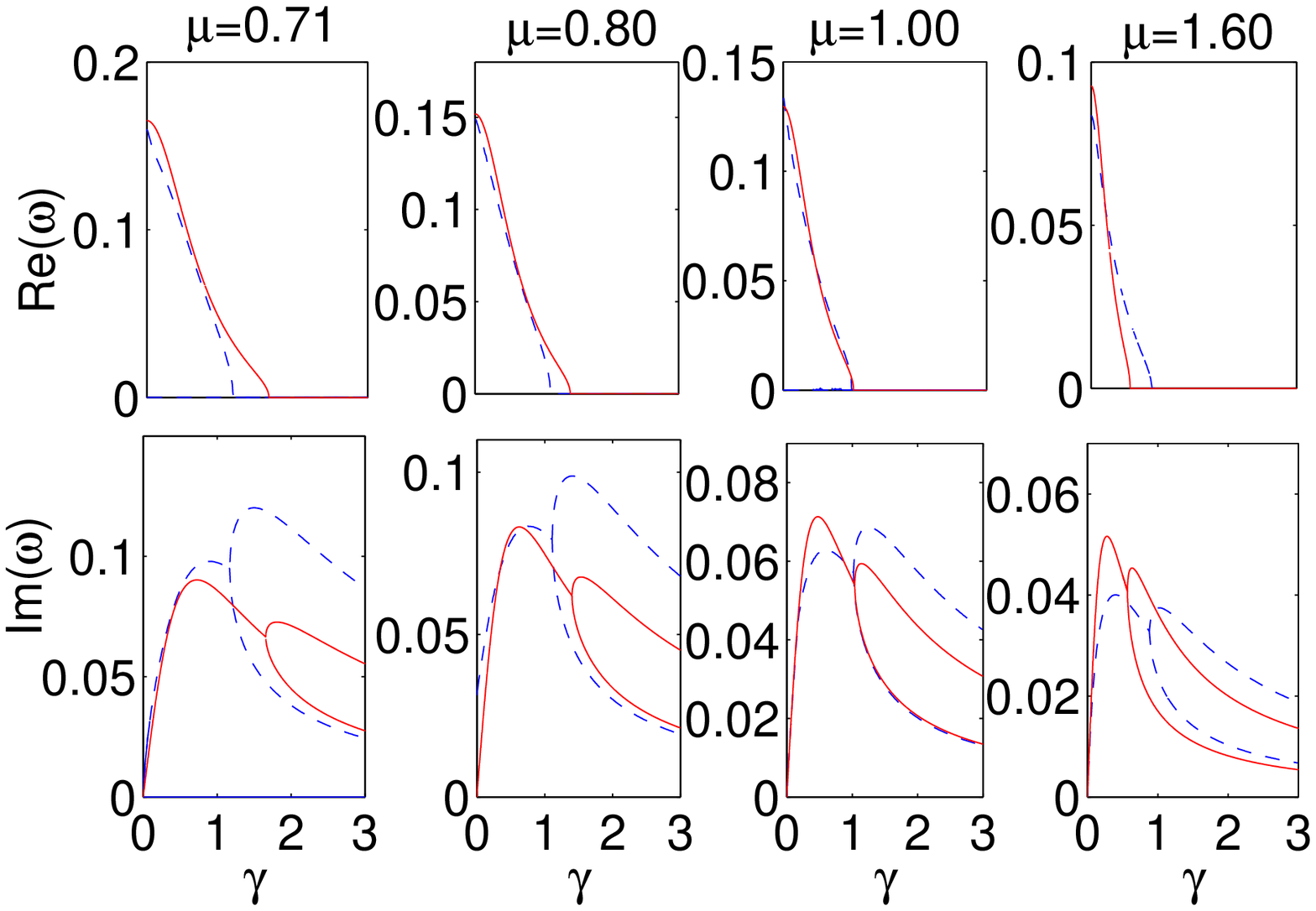}
\caption{
(Color online)
Left Images: The upper four subpanels show the comparison of the imaginary (growth)
part of the
vortex linearization eigenfrequency
associated with the anomalous mode from the BdG numerical analysis (solid blue) with the analytical results $\omega_i$ [see Eq.~(\ref{omega_i})]
(red dashed). The lower four subpanels
show the comparison of the real (oscillatory)
part of the eigenfrequency associated with the anomalous mode from
the BdG numerical analysis (solid blue) with the analytical results
$\omega_r$ [see Eq.~(\ref{omega_r})]
(red dashed).
From the left to the right, the cases shown are
for $\mu=0.4, 0.8, 1, 1.6$, respectively, and $\Omega=0.2$.
Notice the progressively improving agreement as $\mu$ increases
and the Thomas-Fermi regime of vortex particle motion is reached.
Right Images: Shown are the internal mode eigenfrequencies of the vortex dipole state associated with the anomalous mode from
the BdG numerical analysis (dashed blue) against the dissipation term $\gamma$
for $\mu=0.71, 0.80, 1, 1.6$ versus the analytical prediction (solid red),
for $\Omega=0.2$.
}
\label{dissipation_compare_s}
\end{figure*}

Next, turning to numerical computations based on
our analytical approximations, we now examine the linearization
(BdG) analysis around a single vortex (Fig.~\ref{dissipation_compare_s}, left plots),
as well as around a stationary VD (Fig.~\ref{dissipation_compare_s}, right plots).
In the former case, we observe the good agreement of our theoretical prediction
of Eqs.~(\ref{x_dot_revise})-(\ref{y_dot_revise}) in comparison with
the anomalous mode (complex) frequency associated with the single vortex
spiraling outward motion; note that the real part of the relevant eigenfrequency is
associated with the precession around the center, while its imaginary
part with the growing radius of the relevant motion.

As an aside, we note in passing that
for sufficiently large $\gamma$ in the vortex dipole case, there
is a collision of the relevant vortex pair epicyclic internal mode
eigenfrequencies. This, in turn, results in the exponential expulsion
of the VD constituents, a feature that is never possible to observe
for isolated vortices. However, a note of caution should be
added here. The values of $\gamma$ [of $O(1)$] where such
phenomenology arises are roughly
3 orders of magnitude above the physically relevant values at least
for the thermal damping DGPE model of BECs considered herein.
Consequently, such phenomenology is simply of mathematical, rather than
physical relevance.

%
Recall that in the case of the VD, in accordance to what
was shown in Ref.~\cite{middel2}, there is a pair of internal
modes of the two-vortex bound state. One of these is a Goldstone
mode of vanishing frequency associated with the free rotation
of the pair around the trap center. However, the second mode
refers to the epicyclic motion around the vortex
dipole equilibrium observed in Ref.~\cite{pra_11};
this motion is also hinted at by the vortex dynamics
of Ref.~\cite{BPA_10}. It is the latter motion that leads to
the instability (oscillatory or purely exponential, for small
or large $\gamma$, respectively) observed herein.
Finally, it is reassuring to note that in the dipole case,
as in the single vortex case of Fig.~\ref{dissipation_compare_s},
the theoretical approximation for the eigenfrequency becomes
more accurate as the chemical potential $\mu$ gets larger where
the vortices behave more like point particles.

%

%
%



\begin{thebibliography}{99}

\bibitem{fetter}  A. L. Fetter and A. A. Svidzinsky,
J. Phys. Condens. Matter {\bf 13}, R135 (2001).

\bibitem{fetter2} A. L. Fetter,
Rev. Mod. Phys. 81, 647 (2009).

\bibitem{newton_review} P. K. Newton and G. Chamoun,
SIAM Rev. {\bf 51}, 501 (2009).

\bibitem{our_review}
P. G. Kevrekidis, R. Carretero-Gonz{\'a}lez,
D. J. Frant\-zes\-ka\-kis, and I. G. Kevrekidis,
Mod.\ Phys.\ Lett.\ B {\bf 18} (2004) 1481--1505.

\bibitem{rcg:BEC_BOOK} P. G. Kevrekidis, D. J. Frant\-zes\-ka\-kis, and
R. Carretero-Gonz{\'a}lez, {\sl Emergent Nonlinear
Phenomena in Bose-Einstein Condensates: Theory and Experiment}.
Springer Series on Atomic, Optical, and Plasma Physics, Vol.~{\bf 45} (2008).

\bibitem{nvortex_probl}  P. K. Newton,
{\it The N-vortex problem: Analytical techniques},
Springer-Verlag (New York, 2001).

\bibitem{Pismen1999}  L. M. Pismen,
{\it Vortices in Nonlinear Fields} (Clarendon, UK, 1999).

\bibitem{yuri1} Y. S. Kivshar, J. Christou, V. Tikhonenko, B. Luther-Davies,
and L. Pismen,
Opt. Comm. {\bf 152}, 198 (1998).

\bibitem{yuri2} A. Dreischuh, S. Chervenkov, D. Neshev, G. G. Paulus, and H. Walther,
J. Opt. Soc. Am. B {\bf 19}, 550 (2002).

\bibitem{chap01:vort2}
K. W. Madison, F. Chevy, W. Wohlleben, and J. Dalibard,
Phys.\ Rev.\ Lett.\ {\bf 84}, 806 (2000).

\bibitem{chap01:vort3}
S. Inouye, S. Gupta, T. Rosenband, A. P. Chikkatur,
A. G{\"o}rlitz, T. L. Gustavson, A. E. Leanhardt, D. E. Pritchard, and W. Ketterle,
Phys.\ Rev.\ Lett.\ {\bf 87}, 080402 (2001).

\bibitem{foot} E. Hodby, G. Hechenblaikner, S. A. Hopkins,
O. M. Marag{\`o}, and C. J. Foot, Phys. Rev. Lett. {\bf 88}, 010405 (2001).

\bibitem{chap01:latt1}
J. R. Abo-Shaeer, C. Raman, J. M. Vogels, and W. Ketterle,
Science {\bf 292}, 476 (2001).

\bibitem{chap01:latt2}
J. R. Abo-Shaeer, C. Raman, and W. Ketterle,
Phys.\ Rev.\ Lett.\ {\bf 88}, 070409 (2002).

\bibitem{chap01:latt3}
P. Engels, I. Coddington, P. C. Haljan, and E. A. Cornell,
Phys.\ Rev.\ Lett.\ {\bf 89} (2002) 100403.

\bibitem{BPA_KZ}  C. N. Weiler, T. W. Neely, D. R. Scherer, A. S. Bradley, M. J. Davis, and B. P. Anderson,
Nature {\bf 455}, 948 (2008).

\bibitem{kibble} T. W. B. Kibble,
J. Phys. A {\bf 9}, 1387 (1976).

\bibitem{zurek2}  W. H. Zurek,
Nature {\bf 317}, 505 (1985).

\bibitem{freilich} D. V. Freilich, D. M. Bianchi, A. M. Kaufman, T. K. Langin, and D. S. Hall,
Science {\bf 320}, 1182 (2010).

\bibitem{pra_11} S. Middelkamp, P. J. Torres, P. G. Kevrekidis, D. J. Frantzeskakis, R. Carretero-Gonz{\'a}lez, P. Schmelcher, D. V. Freilich, and D. S. Hall
Phys. Rev. A {\bf 84}, 011605 (2011).

\bibitem{BPA_10}  T. W. Neely, E. C. Samson, A. S. Bradley, M. J. Davis, and B. P. Anderson,
Phys. Rev. Lett. {\bf 104}, 160401 (2010).

\bibitem{bagnato} J. A. Seman, E. A. L. Henn, M. Haque, R. F. Shiozaki, E. R. F. Ramos, M. Caracanhas, P. Castilho, C. Castelo Branco, P. E. S. Tavares, F. J. Poveda-Cuevas, G. Roati, K. M. F. Magalhaes, and V. S. Bagnato
Phys. Rev. A {\bf 82}, 033616 (2010).

\bibitem{corot} R. Navarro,
R.\ Carretero-Gonz\'{a}lez,
P. J.\ Torres,
P. G.\ Kevrekidis,
D. J.\ Frantzeskakis,
M. W.\ Ray,
E. Altunta\c{s}, and
D. S.\ Hall, Phys. Rev. Lett. {\bf 110}, 225301 (2013).

\bibitem{Proukakis_Book} N. P. Proukakis, S. A. Gardiner, M. J. Davis, and M. H. Szymanska (Eds.),
{\sl Quantum Gases: Finite Temperature and Non-Equilibrium Dynamics} (Imperial College Press, London, 2013).

\bibitem{shl1} P. O. Fedichev, A. E. Muryshev, and G. V. Shlyapnikov, Phys. Rev. A {\bf 60}, 3220 (1999).

\bibitem{shl2} A. Muryshev, G. V. Shlyapnikov, W. Ertmer, K. Sengstock, and M. Lewenstein,
Phys. Rev. Lett. {\bf 89}, 110401 (2002).

\bibitem{us} S. P. Cockburn, H. E. Nistazakis, T. P. Horikis, P. G. Kevrekidis,
N. P. Proukakis, and D. J. Frantzeskakis, Phys. Rev. Lett. {\bf 104}, 174101 (2010);
{\it ibid.} Phys. Rev. A {\bf 84}, 043640 (2011).

\bibitem{ft1} N. P. Proukakis, N. G. Parker, C. F. Barenghi, and C. S. Adams, Phys. Rev. Lett. {\bf 93}, 130408 (2004).
\bibitem{ft2}
B. Jackson, N. P. Proukakis, and C. F. Barenghi, Phys, Rev. A {\bf 75}, 051601 (2007);
B. Jackson, C. F. Barenghi, and N. P. Proukakis, J. Low Temp. Phys. {\bf 148}, 387 (2007);
\bibitem{ft3}
W. H. Zurek, Phys. Rev. Lett. {\bf 102}, 105702 (2009);
B. Damski, and W. H. Zurek, Phys. Rev. Lett. {\bf 104}, 160404 (2010).
\bibitem{ft4}
A. D. Martin and J. Ruostekoski, Phys. Rev. Lett. {\bf 104}, 194102 (2010);
A. D. Martin and J. Ruostekoski, New J. Phys. {\bf 12}, 055018 (2010).

\bibitem{gk} D. M. Gangardt and A. Kamenev, Phys. Rev. Lett. {\bf 104}, 190402 (2010).

\bibitem{ashton} K. J. Wright and A. S. Bradley, arXiv:1104.2691.

\bibitem{han1} S. Burger, K. Bongs, S. Dettmer, W. Ertmer, K. Sengstock, A. Sanpera,
G. V. Shlyapnikov, and M. Lewenstein, Phys. Rev. Lett. {\bf 83}, 5198 (1999).

\bibitem{han2} K. Bongs, S. Burger, S. Dettmer, D. Hellweg, J. Arlt, W. Ertmer, and K. Sengstock,
C. R.  Acad. Sci. Paris {\bf 2}, 671 (2001).

\bibitem{nist} J. Denschlag, J. E. Simsarian, D. L. Feder, C. W. Clark, L. A. Collins, J. Cubizolles, L. Deng,
E. W. Hagley, K. Helmerson, W. P. Reinhardt, S. L. Rolston, B. I. Schneider, and W. D. Phillips,
Science {\bf 287}, 97 (2000).

\bibitem{bongs} C. Becker, S. Stellmer, P. Soltan-Panahi, S. D\"{o}rscher, M. Baumert,
                E. M. Richter, J. Kronj\"{a}ger, K. Bongs, and K. Sengstock, Nature Physics {\bf 4}, 496 (2008);
                S. Stellmer, C. Becker, P. Soltan-Panahi, E. M. Richter, S. D\"{o}rscher, M. Baumert,
                J. Kronj\"{a}ger, K. Bongs, and K. Sengstock, Phys. Rev. Lett. {\bf 101}, 120406 (2008)

\bibitem{heidelberg} A. Weller, J. P. Ronzheimer, C. Gross, J. Esteve, M. K. Oberthaler, D. J. Frantzeskakis, G. Theocharis, and P. G. Kevrekidis, Phys. Rev. Lett. {\bf 101}, 130401 (2008); G. Theocharis, A. Weller, J. P. Ronzheimer, C. Gross, M. K. Oberthaler, P. G. Kevrekidis, and D. J. Frantzeskakis, Phys. Rev. A, {\bf 81}, 063604 (2010)

\bibitem{ba00} Th. Busch and J. R. Anglin
Phys. Rev. Lett. {\bf 84}, 2298 (2000).

\bibitem{freq} G. Theocharis, P. G. Kevrekidis, M. K. Oberthaler, and D. J. Frantzeskakis,
Phys. Rev. A {\bf 76}, 045601 (2007).

\bibitem{Zwierlein} T. Yefsah, A. T. Sommer, M. J. H. Ku, L. W. Cheuk, W. J. Ji, W. S. Bakr, and M. W. Zwierlein, Nature {\bf 499}, 426 (2013)

\bibitem{djf} D. J. Frantzeskakis, J. Phys. A: Math. Theor. {\bf 43}, 213001 (2010).

\bibitem{lp} L. P. Pitaevskii, Zh. Eksp. Teor. Fiz. {\bf 35}, 408 (1958)
[Sov. Phys. JETP {\bf 35}, 282 (1959)].

\bibitem{penckw} A. A. Penckwitt, R. J. Ballagh, and C. W. Gardiner,
Phys. Rev. Lett. {\bf 89}, 260402 (2002).

\bibitem{npprev} B. Jackson and N. P. Proukakis,
J. Phys. B: At. Mol. Opt. Phys. {\bf 41}, 203002 (2008).

\bibitem{blakie} P. B. Blakie, A. S. Bradley, M. J. Davis, R. J. Ballagh,
and C. W. Gardiner, Adv. Phys. {\bf 57}, 363 (2008).

\bibitem{ZNG_Book}
A. Griffin, T. Nikuni, and E. Zaremba,
{\sl Bose-Condensed Gases at Finite Temperatures} (Cambridge University Press, Cambridge, 2009).

\bibitem{stoof_sgpe}
H. T. C. Stoof, J. Low Temp. Phys. {\bf 114}, 11 (1999); H. T. C. Stoof, and M. J. Bijlsma, J. Low Temp. Phys {\bf 124}, 431 (2001).

\bibitem{stoch} S. P. Cockburn and N. P. Proukakis,
Laser Phys. {\bf 19}, 558 (2009).

\bibitem{dcdss} P. G. Kevrekidis and D. J. Frantzeskakis,
Discr. Cont. Dyn. Sys. S {\bf 4}, 1199 (2011).

\bibitem{njp} V Achilleos, D Yan, P G Kevrekidis, and D J Frantzeskakis,
New J. Phys. {\bf 14}, 055006 (2012).

\bibitem{proukv1} B. Jackson, N. P. Proukakis, C. F. Barenghi, and E. Zaremba, Phys. Rev. A {\bf 79}, 053615 (2009).

\bibitem{proukv2} A. J. Allen, E. Zaremba, C. F. Barenghi, N. P. Proukakis,
Phys. Rev. A {\bf 87}, 013630 (2013).

\bibitem{ourjpb}  S. Middelkamp, P. G. Kevrekidis, D. J. Frantzeskakis, R. Carretero-Gonz{\'a}lez, P. Schmelcher
J. Phys. B {\bf 43}, 155303 (2010).

\bibitem{ashtone} S. J. Rooney, A. S. Bradley, and P. B. Blakie
Phys. Rev. A {\bf 81}, 023630 (2010).

\bibitem{tmwright} T. M. Wright, A. S. Bradley, and R. J. Ballagh, Phys. Rev. A {\bf 80}, 053624 (2009)

\bibitem{spirn} M. Kurzke, C. Melcher, R. Moser, and D. Spirn,
Indiana Univ. J. Math. {\bf 58}, 2597 (2009).

\bibitem{miot} E. Moit, Anal. PDE {\bf 2}, 159 (2009)

\bibitem{stamp}
L. Thompson and P. C. E. Stamp, Phys. Rev. Lett. {\bf 108}, 184501 (2012); T. Cox and P. C. E. Stamp, arXiv:1207.4237 [cond-mat.quant-gas]


\bibitem{stoof} R. A. Duine, B. W. A. Leurs, and H. T. C. Stoof,
Phys. Rev. A {\bf 69}, 053623 (2004).

\bibitem{ashtone2} A. S. Bradley, C. W. Gardiner,
http://xxx.lanl.gov/abs/cond-mat/0509592.



\bibitem{coles} D. E. Pelinovsky and P. G. Kevrekidis
Nonlinearity {\bf 24}, 1271 (2011); see also for the
case of dark solitons M. P. Coles, D. E. Pelinovsky, and P. G. Kevrekidis
Nonlinearity {\bf 23}, 1753 (2010).

\bibitem{footnote1} As this model is of broader mathematical interest, for completeness,
our analysis in Appendix B also extends beyond this regime, even if not directly relevant to cold atoms.


\bibitem{sandst} T. Kapitula, P. G. Kevrekidis, and B. Sandstede,
Phys. D {\bf 195}, 263 (2004).



\bibitem{lundh} E. Lundh and P. Ao,
Phys. Rev. A {\bf 61}, 063612 (2000).

\bibitem{berloff} N. G. Berloff, J. Phys. A {\bf 37}, 1617 (2004).

\bibitem{note} In a very recent work, the motion of a single vortex and of a vortex dipole was explored numerically
in the framework of the stochastic GPE in S. Gautam et al, arXiv:1309.6205.

\bibitem{jerrard} R. L. Jerrard and D. Smets,
arXiv:1301.5213.


\bibitem{busch} S. McEndoo and Th. Busch
Phys. Rev. A {\bf 79}, 053616 (2009).

\bibitem{middel2} S. Middelkamp, P. G. Kevrekidis, D. J. Frantzeskakis, R. Carretero-Gonz{\'a}lez, and P. Schmelcher
Phys. Rev. A {\bf 82}, 013646 (2010).

\bibitem{prouk12} N. G. Parker, A. J. Allen, C. F. Barenghi, and N. P. Proukakis,
Phys. Rev. A {\bf 86}, 013631 (2012).

\bibitem{allen_soliton} A. J. Allen, D. P. Jackson, C. F. Barenghi, and N. P. Proukakis,
Phys. Rev. A {\bf 83}, 013613 (2011).

\bibitem{footnote2}
It should be noted that in this case the condensate is sufficiently
wide that the homogeneous case value of $B=2$ has been utilized.

\bibitem{bpa_vr} B. P. Anderson, P. C. Haljan, C. A. Regal, D. L. Feder, L. A. Collins, C. W. Clark, and E. A. Cornell,
Phys. Rev. Lett. {\bf 86}, 2926 (2001).

\bibitem{hau} N. S. Ginsberg, J. Brand, and L. V. Hau,
Phys. Rev. Lett. {\bf 94}, 040403 (2005).

\bibitem{komineas} S. Komineas,
Eur. Phys. J. Spec. Top. {\bf 147}, 133 (2007).

\end{thebibliography}
\end{document}